\documentclass{aa}  %

\usepackage{graphicx}
\usepackage{txfonts}
\usepackage{xcolor}

\begin{document}

   \title{An analysis of the J-method from the perspective of the AGB evolution}


   \author{C. Gavetti\inst{1,2}, P. Ventura\inst{2}, F. Dell'Agli\inst{2}, M. Correnti\inst{2,3},
             F. La Franca\inst{1}}

   \institute{Dipartimento di Matematica e Fisica, Università degli Studi Roma Tre, 
              via della Vasca Navale 84, 00100, Roma, Italy \and
              INAF, Observatory of Rome, Via Frascati 33, 00077 Monte Porzio Catone (RM), Italy \and
              ASI-Space Science Data Center, Via del Politecnico, I-00133, Rome, Italy
              }

   \date{Received September 15, 1996; accepted March 16, 1997}


 \abstract
   {The JAGB method has been proposed over the last years as a possible distance
    indicator for the galaxies in the Local Group and possibly beyond. The nature of
    the stars populating the J region, as also the conditions
    on the star formation history and on the structural properties of the galaxies
    for the straight application of this method need still to be investigated.}
   {We study the populations of the J region of the colour-magnitude
   $\rm (J-K,J)$ plane of the Large and Small Magellanic Cloud (LMC and SMC, respectively), 
   to relate the shape of the J luminosity function (JLF) to the details of the
   formation histories of the two galaxies, in the attempt of distinguishing the 
   general aspects of the JLF to those more sensitive to the stellar
   population of the specific galaxy considered.}
   {We use a population synthesis approach, based on the combined results from stellar
    evolution and dust formation modelling, to find the expected distribution of the stars
    within the J region, and compare it with that derived from the observations of LMC 
    and SMC stars. Some physical
    assumptions, mostly related to the modelling of the red giant branch and
    asymptotic giant branch phases of stars, are tuned, until satisfactory agreement between 
    the expectations from synthetic modelling and the observational evidence is
    reached.}
   {The sources observed within the J region are identified with stars that have
   recently reached the C-star stage, and have not yet accumulated the extremely large
   amounts of carbon required to make the evolutionary track to evolve off the J region.
   Generally speaking, $\rm 2-3~M_{\odot}$ stars stay longer within the J region,
   while lower mass objects evolve there for at most a couple of inter-pulse phases.
   The analysis of the JLF of the LMC, peaked at the J magnitudes expected for these
   stars, confirm this understanding. In the SMC the distribution of the J fluxes is
   shifted to higher J magnitudes when compared to the LMC, which we interpret as the
   signature of an average older population, with smaller mass progenitors.
    }
   {}

   \keywords{stars: AGB and post-AGB -- stars: abundances -- stars: evolution -- stars: mass-loss
               }

   \titlerunning{The J Region Populations of the Magellanic Clouds}
   \authorrunning{Gavetti et al.}
   \maketitle
%

\section{Introduction}
During the past decades, increasing attention has been devoted to the asymptotic giant branch (AGB), the evolutionary phase experienced by all stars with initial mass between $\rm 1$ and $\rm 8~M_{\odot}$, which begins after the
consumption of central helium, and extends until the external envelope of
the star is lost. AGB stars are often split into different sub-classes,
such as those that haven't experienced any thermal pulse yet (early-AGB),
those enriched in carbon in the surface regions (carbon stars), or those
surrounded by dust. An example of the distinction among AGB stars is
found in \citet{weinberg01}. During the AGB phase the stars lose their envelopes, enriching the surrounding medium with chemically processed gas \citep{ciaki20,romano22}. These ejecta play a key role in the production of carbon and nitrogen \citep{fiorenzo}, in shaping the chemical patterns of Local Group star-forming regions, and possibly in the origin of multiple stellar populations in globular clusters \citep{ventura01}.

AGB stars are also among the most efficient dust producers, as their circumstellar envelopes provide ideal conditions for grain condensation \citep{gase85}. Together with supernovae, they dominate the cosmic dust budget, although their relative contributions remain debated \citep{raffa23}. Understanding AGB dust formation is thus essential both for computing stellar dust yields and for interpreting IR observations, given the reprocessing of stellar radiation by circumstellar dust.

Modern AGB models now include self-consistent treatments of dust formation in stellar winds \citep{ventura12, ventura14, nanni13, nanni14}, following the approach of the Heidelberg group \citep{fg01, fg02, fg06}, which predict dust composition, quantity and production rates. These models have been widely applied to estimate the AGB dust budget in nearby galaxies \citep{raffa14} and to interpret evolved stellar populations in the Magellanic Clouds (MCs) and other Local Group systems \citep{flavia14b, flavia15a, flavia15b, nanni16, nanni19, flavia16, flavia18, flavia19, cla}.

The capability to investigate the evolved stellar populations of galaxies is becoming increasingly relevant in the James Webb Space Telescope (JWST) era. Due to its unprecedented infrared sensitivity, JWST enables the detection of resolved stellar populations in galaxies well beyond the Local Group and provides an exceptional tool for studying AGB stars \citep{correnti25, bortolini25}. For a significant fraction of these systems, AGB stars represent important tracers for reconstructing their star formation histories (SFHs), due to their age sensitivity in the NIR \citep{Lee24b,bortolini24}.

A further motivation to study the structure, evolution and dust formation of AGB stars is their potential use as distance indicators. This idea originates from \citet{nikolaev00}, who identified a region in the $\rm (J-K_s, K_s)$ plane of the LMC - later called "J region" - dominated by carbon stars and proposed them as standard candles to trace the 3D structure of the LMC. 
Subsequent works \citep{madore20, freedman20} extended this approach to nearby galaxies, coining the JAGB method, and found that stars within $\rm 1.3 < J-K < 2.0$ mag have a mean $\rm M_J \sim -6.20$ mag. However, \citet{ripoche20} reported systematic differences between LMC, SMC and Milky Way, suggesting a possible metallicity dependence.
More recently, \citet{magnus24} refined the calibration using Gaia data, selecting a clean carbon star sample within $\rm 1.5 < J-K < 2.0$ mag and a 1.2 mag wide J window. They confirmed a mean $\rm M_J$ of $\sim -6.25$ mag for the LMC and $\sim -6.18$ mag for the SMC. 

The results discussed above highlight the need for a precise calibration of the JAGB method to enable its use for measuring distances to more distant galaxies. Indeed, JAGB stars are, on average, significantly brighter than the tip of the red giant branch (TRGB), allowing distance determinations well beyond the TRGB limit. Compared to Cepheids, the JAGB method requires only a single epoch of observations and can be applied to galaxies hosting stellar populations aged between $\sim 200$ Myr and 1 Gyr, whereas Cepheids are confined to the disks of spiral and irregular systems. These advantages, together with the advent of JWST, have motivated the community to apply this technique to an increasing number of galaxies to refine the determination of the Hubble constant \citep{Lee24a,Li25,freedman25}.

Against this background, we decided to start a new research
project, designed to understand if and under which conditions the
JAGB method can be safely used to determine the distance of galaxies,
based on the study of the evolution of the positions
of the stars on the CMD, as they evolve through the AGB. To this aim
we consider AGB models of different mass and chemical composition,
where dust formation is taken into account, to allow  the
variation of the spectral energy distribution (SED) to be followed, 
then the determination of the IR colours and
magnitudes. We concentrate on the evolutionary phases during which 
the stars evolve across the J region, defined according to the
recommendations by \citet{magnus24}, to clarify the following points: 
a) which stars enter the J region?; b) which is the duration of
the stay of the stars within the J region?; c) which is the 
chemical composition, the dust mineralogy and the dust production rate (DPR) of the
stars during these evolutionary phases? Answering these questions is 
the sine qua non condition for a general application of the J method 
to measure distances of galaxies.

In this first work we describe the main factors affecting the
crossing and the time of stay of AGB stars in the J region, 
in relation to the mass and the formation epoch of the stars.
We start by analysing the LMC and the SMC, since for these two 
galaxies both the SFH and the age-metallicity relationship (AMR) 
are robustly known, which makes
the comparison between theoretical modelling and observations
easier. We compare the J luminosity function of the stars in 
the J region of the CMDs of the LMC and SMC published in \citet{magnus24},
with results based on a population synthesis approach, which 
relies on the modelling of the AGB evolution and the dust
formation process. 
The goal of the present analysis, other than
confirming the possibility of reproducing the observed luminosity 
functions, is the characterisation of the LMC and SMC stars 
nowadays populating the J region, in relation to the mass, chemical
composition and formation epoch of the progenitors. 

This work is to be considered as a first step towards the
comprehension of the conditions under which the J method can be
reliably applied as a distance indicator, an argument that will
be deepened in the following investigations on this topic, where
all the galaxies for which the JLF has been derived will be
considered, and a general overview of the results obtained
with various SFH and AMR will be presented.

The structure of the paper is as follows: the numerical and physical
ingredients used to model the AGB evolution and the dust formation process,
and the techniques adopted in the population synthesis approach, are
described in section \ref{input}; the main physical and chemical
properties of AGB stars, and the conditions and timing of their
crossing the J region of the colour-magnitude $\rm (J-K,J)$ plane, are
discussed in section \ref{agb}; section \ref{lmc} is devoted to
the application of the population synthesis approach to study
the stellar population of the J region of the LMC, to test the
possibility to reproduce the results from \citet{magnus24};
a comparative analysis between the populations of the J regions
of the LMC and SMC, in relation to the difference in the SFH of
the two galaxies, is addressed in section \ref{smc}; finally,
the conclusions are given in section \ref{concl}.

\section{Physical and numerical input}
\label{input}
The analysis presented in this work is based on the
comparison between the observed J luminosity function of the stars 
populating the J region of the LMC and the SMC, selected on the
basis of the indications given in \citet{magnus24}, with the results 
obtained by a population synthesis approach. In the latter 
we run numerical simulations, based on the SFH and the 
AMR of the galaxy considered, to calculate the number of 
stars nowadays evolving through the JAGB, and the mass and metallicity of 
the progenitors. For the LMC we adopted the SFH derived by \citet{mazzi21}, 
combined with the AMR by \citet{carrera08}, while for the SMC 
we followed the SFH and the AMR given by \citet{rubele18}.

\begin{figure*}
\vskip-40pt
\begin{minipage}{0.46\textwidth}
\resizebox{1.\hsize}{!}{\includegraphics{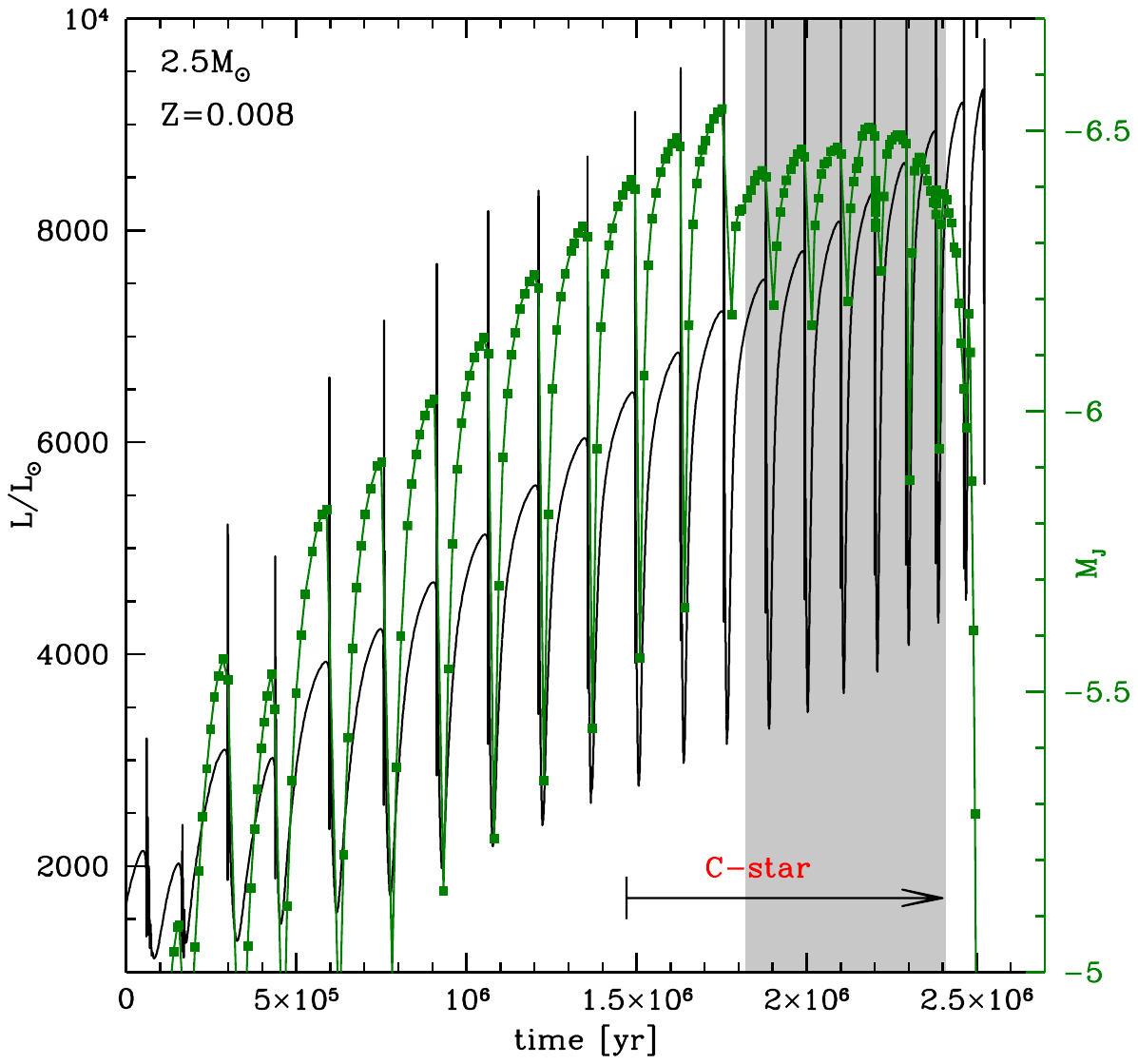}}
\end{minipage}
\begin{minipage}{0.46\textwidth}
\resizebox{1.\hsize}{!}{\includegraphics{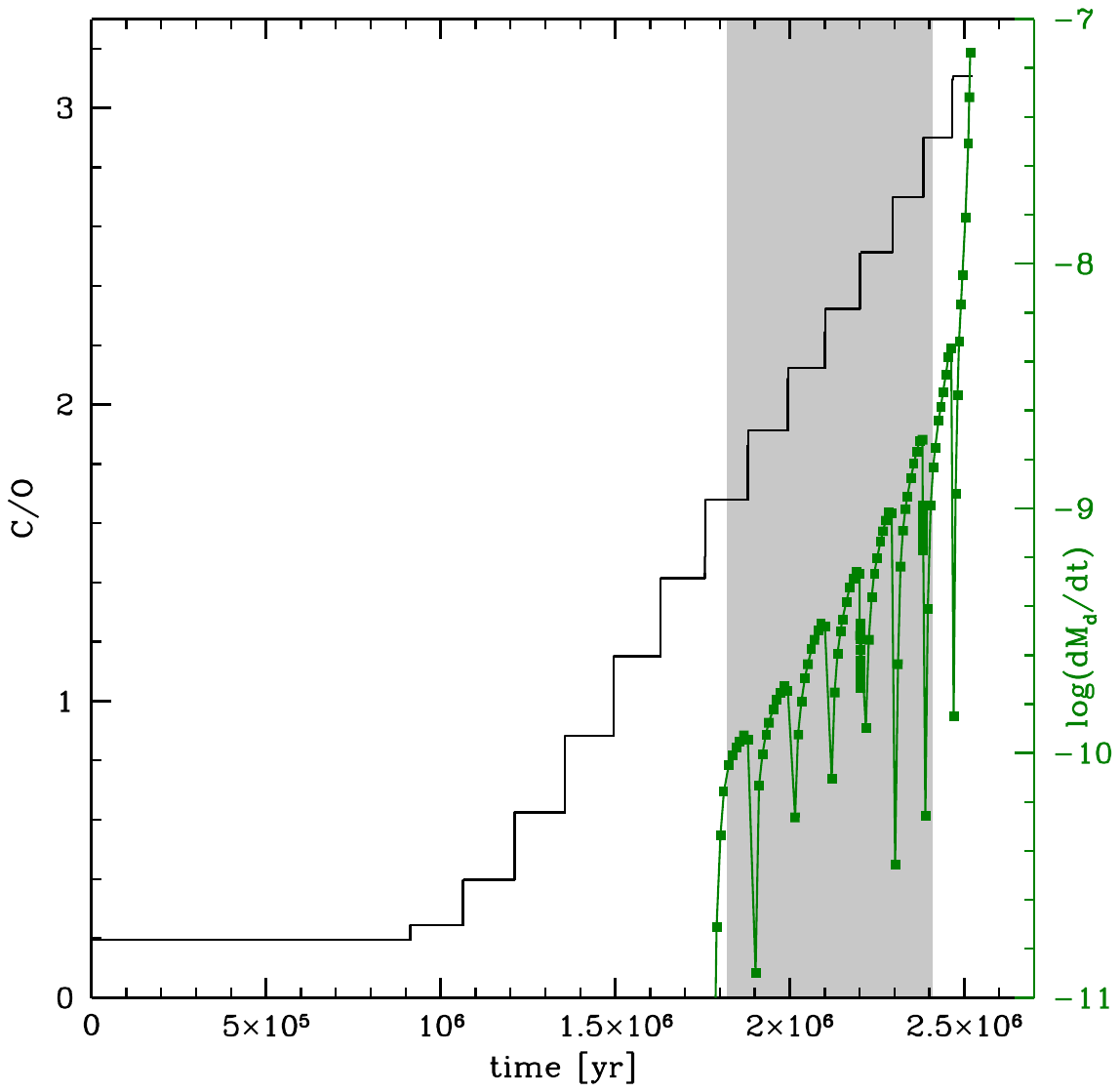}}
\end{minipage}
\vskip-60pt
\caption{{\bf {Left}}: Time variation of the luminosity (black line,
scale on the left) and the J magnitude (green, scale on the right)
of a model star of initial mass $\rm 2.5~M_{\odot}$ and metallicity 
$\rm Z=0.008$. The black arrow indicates the start of the C-star phase.
{\bf Right:} Time evolution of the surface $\rm C/O$ and the DPR of the
same model star shown in the left panel. Times are counted since the
beginning of the TP-AGB phase. Grey-shaded regions indicate the
phase during which $\rm 1.5 < J-K < 2$ mag.
} 
\label{f25}
\end{figure*} 

Finding the synthetic distribution of the stars across the CMD demands
the knowledge of the evolutionary tracks of the stars evolving 
across the AGB and the time variation of the position along the
track. This is obtained by modelling the evolution of stars of different
mass and metallicity, combined with the description of the dust formation, 
which is required to account for the modification of the SED
due to the reprocessing of the radiation by dust grains.
For this work we used extant evolutionary sequences previously published 
by our group, of metallicity $Z=0.001$ \citep{ventura14}, $Z=0.004$ and 
$Z=0.008$ \citep{marini21}. The above computations were recently extended 
until the start of the white dwarf cooling, as described in \citet{devika23}. 
For the stars of initial mass $\rm M \leq 2~M_{\odot}$,
the computations are first evolved from the pre-MS through the core 
hydrogen burning and the red giant branch (RGB) phase, until the TRGB, 
when the helium flash takes place.
The simulations are then re-started from the 
quiescent core helium burning phase, based on the core masses reached at the TRGB. 
These stars lose a significant fraction of the mass of their envelope 
during the evolution along the red giant branch (RGB), thus the mass with which
they undergo core helium burning is smaller than the initial mass. We assumed
that all the stars of mass below $\rm 1.5~M_{\odot}$ lose $\rm 0.2~M_{\odot}$
during the ascent along the RGB. In the following we will refer to the 
mass of these stars at the start of the core helium burning phase, keeping in
mind that the mass of the progenitors was $\rm 0.2~M_{\odot}$ higher.
We will test how this choice affects the
results obtained.

Dust formation in the wind is described by following the formalism proposed by 
\citet{fg06}, in the modality extensively described in \citet{ventura12}. The
application of the modelling of dust formation to the results from stellar
evolution leads to the determination of the variation of the dust composition
and the DPR of the stars as they evolve along the AGB. 
The final step to build the evolutionary tracks of the stars is the determination 
of the variation of the SED, which, in turn, is used to find the colours and
magnitudes in the selected filters, via convolution of the SED with
the corresponding transmission curves. This is done by means
of the DUSTY code \citep{nenkova99}, which uses as input the results from
the modelling of dust formation. This work is based on the sequence of
synthetic SEDs recently used by \citet{cla}.

The evolutionary sequences of the stars of different initial
mass and metallicity are available at the CDS. For $\rm M > 2~M_{\odot}$
stars the masses refer to the initial values, with which the stars formed;
for the lower mass counterparts the masses are taken at the start of the
core helium burning phase. For each model
star we list the variation during the AGB phase of the most
relevant physical quantities and of the JHK magnitudes.

\section{The AGB evolution across the J region}
\label{agb}
\subsection{The main physical aspects of the AGB evolution}
The AGB evolution is driven by the gradual growth of the core mass
($\rm M_C$), under the action of the H-burning shell, which for the
majority of the AGB lifetime is the unique nuclearly active source
in the star \citep{karakas14}. Periodically, helium ignition occurs in 
a thin shell lying above the degenerate core made up of carbon:
because these nuclear episodes occur in conditions of thermal 
instability \citep{sch},
the common terminology used to refer to them is thermal pulse (TP). The rise
in the core mass favours the increase in the luminosity, given the
approximately linear relationship connecting the two quantities,
originally derived by \citet{pacz}, to which all AGB stars
obey, with some exceptions that will be discussed later in this
section. This behaviour can be seen in the left panels of 
Fig.~\ref{f25} and Fig.~\ref{f40}, which report the AGB evolution of 
model stars of initial mass $\rm 2.5~M_{\odot}$ and $\rm 4~M_{\odot}$,
respectively. In the $\rm 2.5~M_{\odot}$ case the luminosity increases 
during the AGB phase from $\rm \sim 2\times 10^3~L_{\odot}$ to 
$\rm \sim 9\times 10^3~L_{\odot}$,
while the core mass changes from $\rm 0.52~M_{\odot}$ to $\rm 0.62~M_{\odot}$.
As for the $\rm 4~M_{\odot}$ model star, the luminosity varies from
$\rm \sim 1.2\times 10^4~L_{\odot}$ to $\rm \sim 3.5\times 10^4~L_{\odot}$, 
with the core mass increasing from $\rm 0.82~M_{\odot}$, at the occurrence 
of the first TP, to $\rm 0.88~M_{\odot}$, at the end of the AGB.

\subsection{The surface chemical composition of AGB stars}
\begin{figure*}
\vskip-40pt
\begin{minipage}{0.46\textwidth}
\resizebox{1.\hsize}{!}{\includegraphics{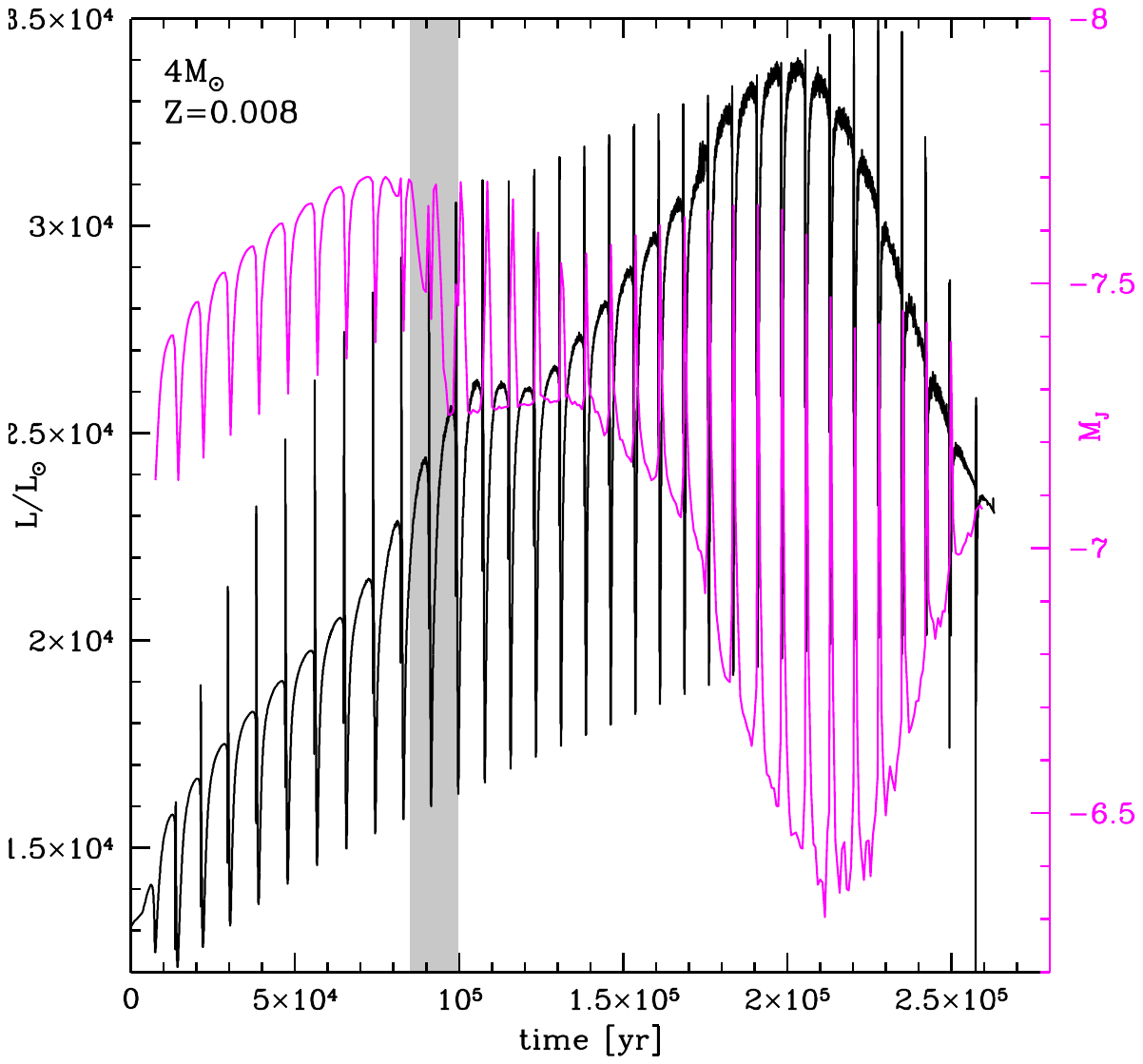}}
\end{minipage}
\begin{minipage}{0.46\textwidth}
\resizebox{1.\hsize}{!}{\includegraphics{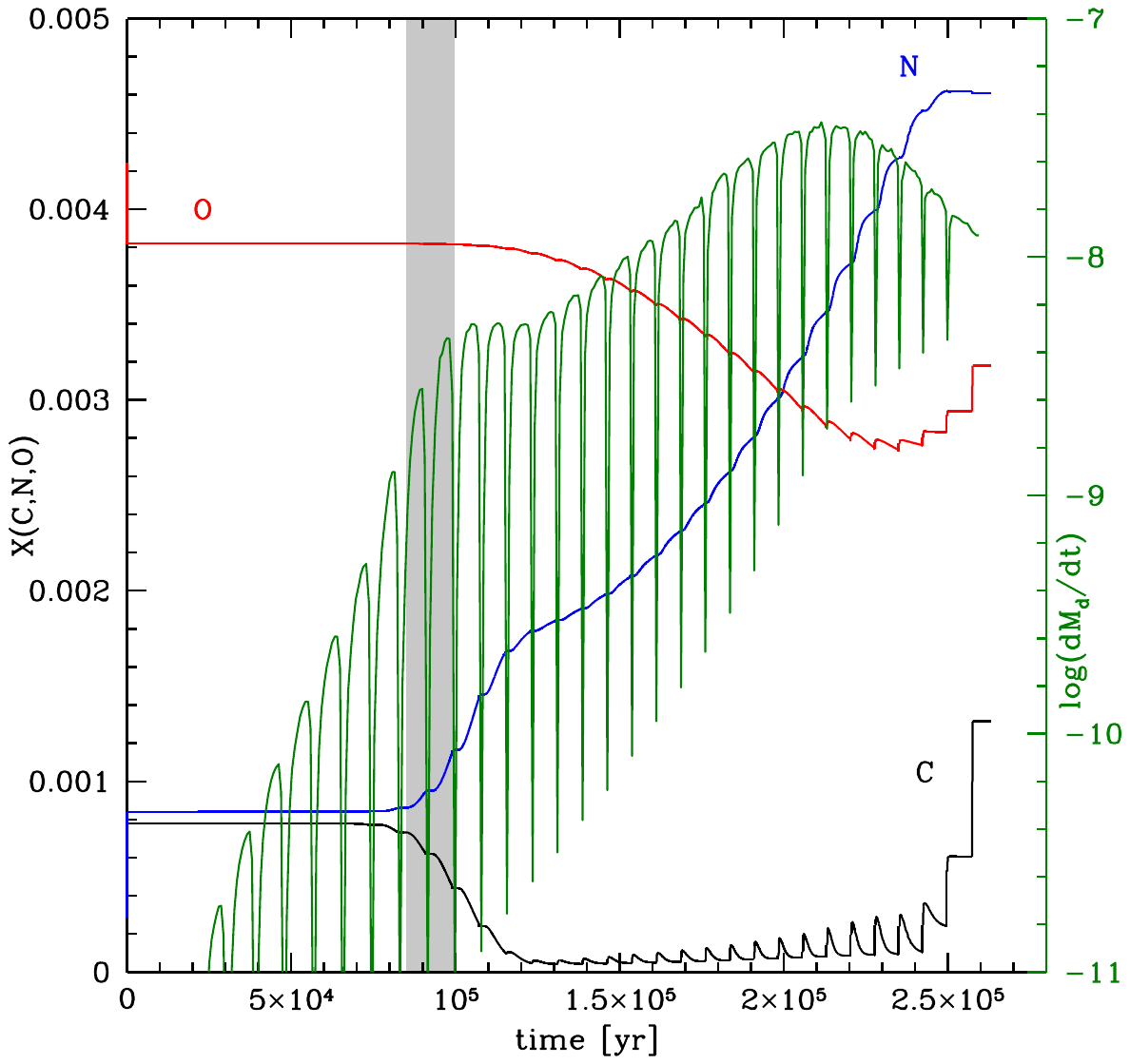}}
\end{minipage}
\vskip-60pt
\caption{{\bf Left:} Time variation of the luminosity (black line,
scale on the left) and the J magnitude (magenta, scale on the right)
of a model star of initial mass $\rm 4~M_{\odot}$ and metallicity 
$\rm Z=0.008$. {\bf Right:} Time evolution of the surface carbon (black line),
nitrogen (blue), oxygen (red) (black scale on the right) and the DPR (green scale on the right) 
of the same model star shown in the left panel. Times are counted since the 
beginning of the TP-AGB phase. Grey-shaded regions indicate the phase during 
which $\rm 1.5 < J-K < 2$ mag.
} 
\label{f40}
\end{figure*}

The surface chemical composition of AGB stars can be altered by two
mechanisms, which leave different chemical imprinting. The third
dredge-up (TDU) takes place during the evolutionary phases following
the ignition of each TP, when the H-burning shell is temporarily extinguished and the surface convection penetrates inwards until
reaching regions of the star previously processed by helium nucleosynthesis
\citep{iben74}. The primary effect of the TDU is the increase in the
surface abundance of carbon, which can lead to the formation of
a carbon star. The alternative mechanism able to change the surface
chemical composition of AGB stars is hot bottom burning (HBB), which
consists in the activation of proton capture nucleosynthesis at the base
of the convective envelope, when the temperatures in those regions of
the star reach (and exceed) 30 MK \citep{sack}. The ignition of HBB
leads to a fast increase in the stellar luminosity \citep{ventura05},
with significant deviations from the linear $\rm M_C-L$ trend found
by \citet{pacz}. The impact of HBB on the behaviour of the luminosity
can be seen in the left panel of Fig.~\ref{f40}, which shows that the
luminosity of the $\rm 4~M_{\odot}$ model star first increases, 
until $\rm \sim 3.5\times 10^4~L_{\odot}$,
then decreases during the final evolutionary phases, because HBB
is progressively turned off by the gradual loss of the envelope. This
behaviour, typical of the stars experiencing HBB, is not seen 
in the time variation of the luminosity of
the $\rm 2.5~M_{\odot}$ model star shown in Fig.~\ref{f25}, because the
ignition of HBB requires core masses in excess of $\rm 0.8~M_{\odot}$
\citep{ventura13}, which are reached only by stars of initial mass above 
$\rm \sim 3~M_{\odot}$. On the chemical side, HBB changes the surface
chemistry of the star according to the equilibria of the p-capture
nucleosynthesis activated. While the degree of the nucleosynthesis
experienced is extremely sensitive to the metallicity of the star
\citep{flavia18}, the depletion of the surface carbon and the parallel
synthesis of nitrogen take place in all the cases when HBB is activated.
The main effects of TDU and HBB can be seen in the right panels
of Fig.~\ref{f25} and \ref{f40}. In the former we note the gradual
rise in the surface $\rm C/O$ ratio, which under the effects of repeated 
TDU episodes grows until a final value around 3. After about half of 
the AGB lifetime, the surface carbon exceeds oxygen  and the 
star reaches the C-star stage (this is indicated by the horizontal arrow 
in the lower part of the figure). In the right panel of Fig.~\ref{f40}
we see the effects of HBB in the decrease in the surface carbon and
the rise in the nitrogen abundance, which start $\sim 10^5$ yr
after the beginning of the AGB phase. During the very final evolutionary
phases the surface carbon increases under the action of TDU, while
HBB is turned off.

\subsection{The position of AGB stars on the $(J-K, K)$ plane}
\label{gruppi}
\citet{ventura22} divided AGB stars in three different groups, according
to the modality with which the surface chemistry changes, and
consequently on the chemistry and the quantity of dust produced in their
wind, and then on the evolution of the SED: I) $\rm M<1~M_{\odot}$ stars,
whose chemical composition is substantially unchanged during the
AGB evolution, because the core mass is too low for the ignition of HBB,
and they lose the external mantle, of a few tenths of solar masses,
before repeated TDU events can rise the surface carbon significantly;
II) $\rm 1~M_{\odot} < M < 3~M_{\odot}$ stars, which become carbon stars,
under the effects of TDU; III) massive AGBs descending from $\rm M>3~M_{\odot}$
progenitors, whose surface chemistry mainly reflects the effects of 
HBB\footnote{The mass limits are slightly dependent on the metallicity.
As discussed in \citet{devika23}, the lower threshold mass required to
reach the C-star stage descreases for lower Z's, because the lower oxygen
content eases the formation of carbon stars in metal-poor enviroments.
The lower limit in mass for the ignition of HBB is lower the lower Z,
given the hotter temperatures at the base of the envelope of 
metal-poor, massive AGBs \citep{flavia18}.}.

\begin{figure}
\vskip-40pt
\centering
\begin{minipage}{0.5\textwidth}
\resizebox{1.\hsize}{!}{\includegraphics{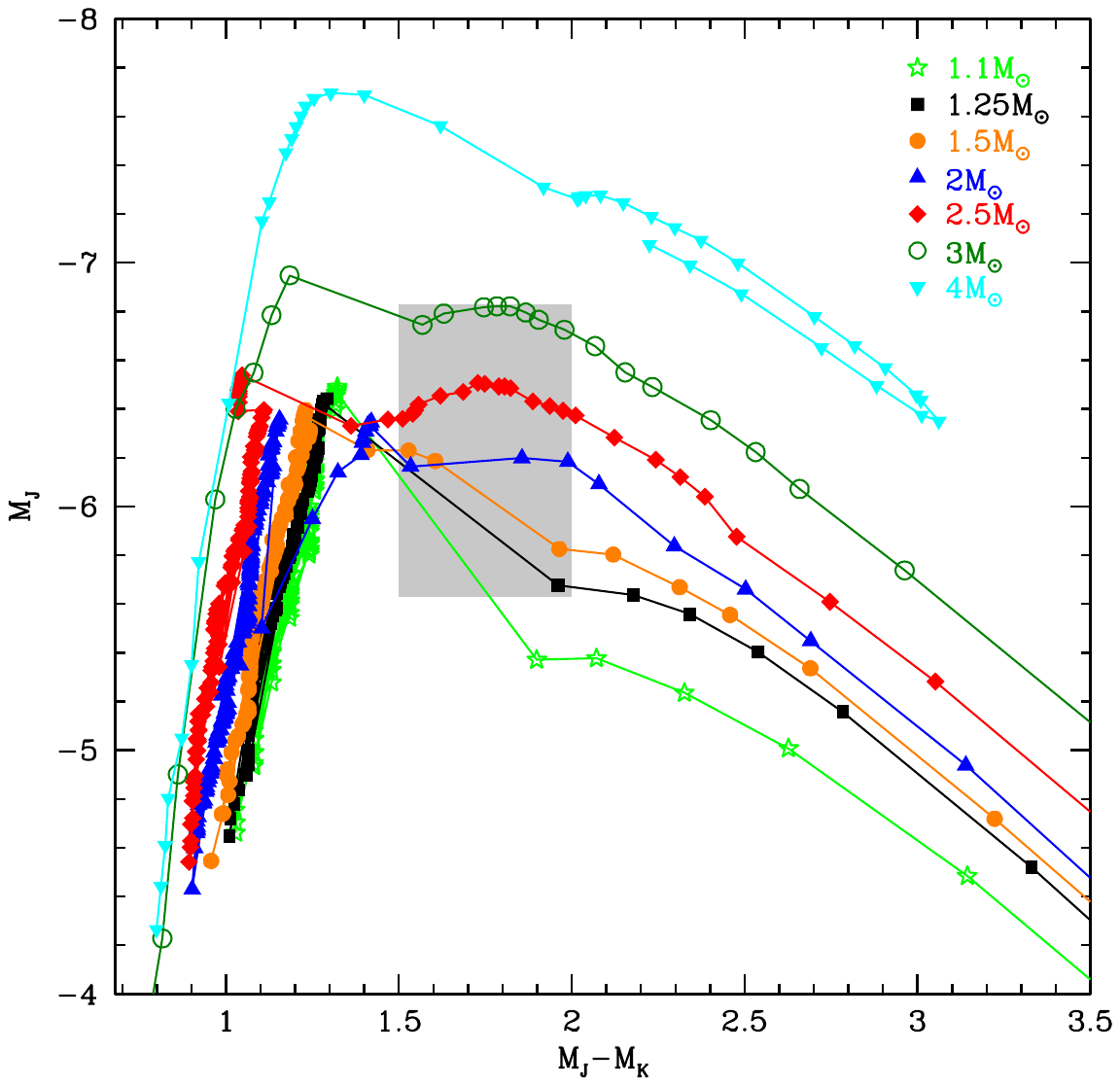}}
\end{minipage}
\vskip-70pt
\caption{Evolutionary tracks of model stars of metallicity
$\rm Z=8\times 10^{-3}$ on the (J-K, J) colour-magnitude 
diagram. The different points along the tracks 
refer to some selected
evolutionary stages taken during the AGB phase, chosen in
order to well represent the excursion of the tracks on the
observational plane. The grey shaded region indicate the box chosen by
\citet{magnus24} for the J region. The masses given for
$\rm M<1.5~M_{\odot}$ refer to the values attained at the TRGB.} 
\label{ftracce}
\end{figure}

Group I) is not of interest for the present investigation, 
because they evolve on the blue side of the CMD, never entering the
J region of the plane. The evolution of the stars belonging to 
group II) is characterised by the gradual increase in the surface 
carbon, which proceeds until the almost complete loss of the envelope.
A prototype of the evolution of these objects is shown in Fig.~\ref{f25}.
The variation of the DPR shown in the right panel of Fig.~\ref{f25}
indicates that dust production is negligible during the O-rich phase,
while it becomes higher and higher as carbon is accumulated in the
surface regions, until reaching values of the order of 
$\rm 10^{-7}~M_{\odot}/$yr during the final part of the AGB evolution.
The increase in the surface carbon enhances the efficiency of the
dust production mechanism for two reasons, related to the thermal
reaction of the external regions of the star to the transition from the
O-rich to the C-star phase and to the higher number of C-bearing
molecules in the external mantle. The first motivation is that 
the formation of carbon stars was shown to trigger the enhancement of
the surface molecular opacities \citep{marigo02}, thus leading to the 
expansion and cooling of the external layers of the star \citep{vm09, vm10},
lower surface gravities, then higher mass loss rates. These conditions
turn extremely favourable to the condensation of gaseous molecules into
solid grains, because the vaporisation process is partly inhibited by the
low temperatures, and the larger densities make a higher number of
gaseous molecules available to condense into dust grains. A further reason
why the higher surface carbon content favours dust production is that
the amount of carbon dust formed depends on the carbon excess with
respect to oxygen \citep{fg06}, thus the number of molecules available
to form carbon dust is higher the higher the surface carbon. 
The formation of carbon dust favours the gradual shift of the SED
towards the near-IR and then the mid-IR spectral region. 

Fig.~\ref{ftracce} shows the evolutionary tracks of stars of different
initial mass and metallicity $\rm Z=8\times 10^{-3}$ in the CMD.
The minimum mass considered in the plot is $\rm 1.1~M_{\odot}$ (which 
corresponds to a progenitor's mass of $\rm 1.3~M_{\odot}$), because 
the tracks of lower mass stars develop along the blue side of the plane,
never turning to the red, thus not crossing the J region, identified with
a grey box in the figure. All the evolutionary tracks reported in
Fig.~\ref{ftracce} correspond to stars belonging to the group II),
but the cyan line, corresponding to the $\rm 4~M_{\odot}$ model star. 
The tracks evolve to the red side of the plane and enter the J region 
soon after becoming carbon stars. 

In the left panel of Fig.~\ref{f25} we note the $\sim 0.15$ mag 
depression in $\rm M_J$ taking place 
shortly after the transition to C-star, despite the luminosity of the star
increases during the same period: this indicates that the peak of the SED
progressively moves to the $\rm \lambda > 1~\mu m$ spectral region. In the
specific $\rm 2.5~M_{\odot}$ case discussed here the star evolves within
the J region of the CMD with $\rm 1.5 < J-K < 2.0$ mag, where it stays 
for $\sim 6\times 10^4$ yr,
during 6 inter-pulse phases following the transition from M type to C-star.
The evolutionary track of the $\rm 2.5~M_{\odot}$ model star discussed here
is shown in red in Fig.~\ref{ftracce}.
During the very final AGB phases, when the surface $\rm C/O$ exceeds 2.5,
the formation of notable quantities of carbon dust favours a significant
reprocessing of the radiation released from the photopshere of the star,
and the shift of the SED to the mid-IR spectral region. Under these conditions
the flux in the J band is almost negligible, so that the evolutionary track 
moves to the lower far red side of the CMD. We see in Fig.~\ref{ftracce} that
this behaviour is common to all the stars that become carbon stars, which
during the final AGB phases populate the K region of plane, according to
the nomencalture adopted by \citet{weinberg01} to identify carbon dust-enshrouded
AGB stars on the observational plane.
The values of $\rm M_J$ attained by the star during these advanced
evolutionary phases derived from the synthetic modelling are to be taken
with some caution; however, this is not an issue for the analysis done here, 
because these stars do not fall into the J region of the CMD. 

These results indicate that $\rm 1-3~M_{\odot}$ stars belonging to the 
group II) introduced earlier in this section evolve into the J region of 
the CMD when they undergo a transition phase, which starts after 
becoming carbon stars. The beginning of this phase occurs when the
carbonaceous dust in the wind forms with sufficiently large rates, so 
that the SED is shifted towards the near IR, and $\rm J-K > 1.5$ mag. 
The conclusion of this transition time takes place when the excess of 
carbon with respect to oxygen becomes so large that the dust forms in
the wind in quantities large enough to lead to the  
$\rm J-K > 2.0$ mag condition. The stars in the J region of the CMD
must have accumulated in the surface regions an amount of carbon 
that corresponds to a specific range of $\rm C/O$, which in the 
$\rm 2.5~M_{\odot}$ case discussed here is $\rm 1.7 < C/O < 2.8$.
We will see that this result cannot be fully generalized, as the
exact $\rm C/O$ range for the $\rm 1.5 < J-K < 2.0$ mag condition
is sensitive to the mass and the metallicity of the star considered.

\begin{figure*}
\vskip-40pt
\begin{minipage}{0.46\textwidth}
\resizebox{1.\hsize}{!}{\includegraphics{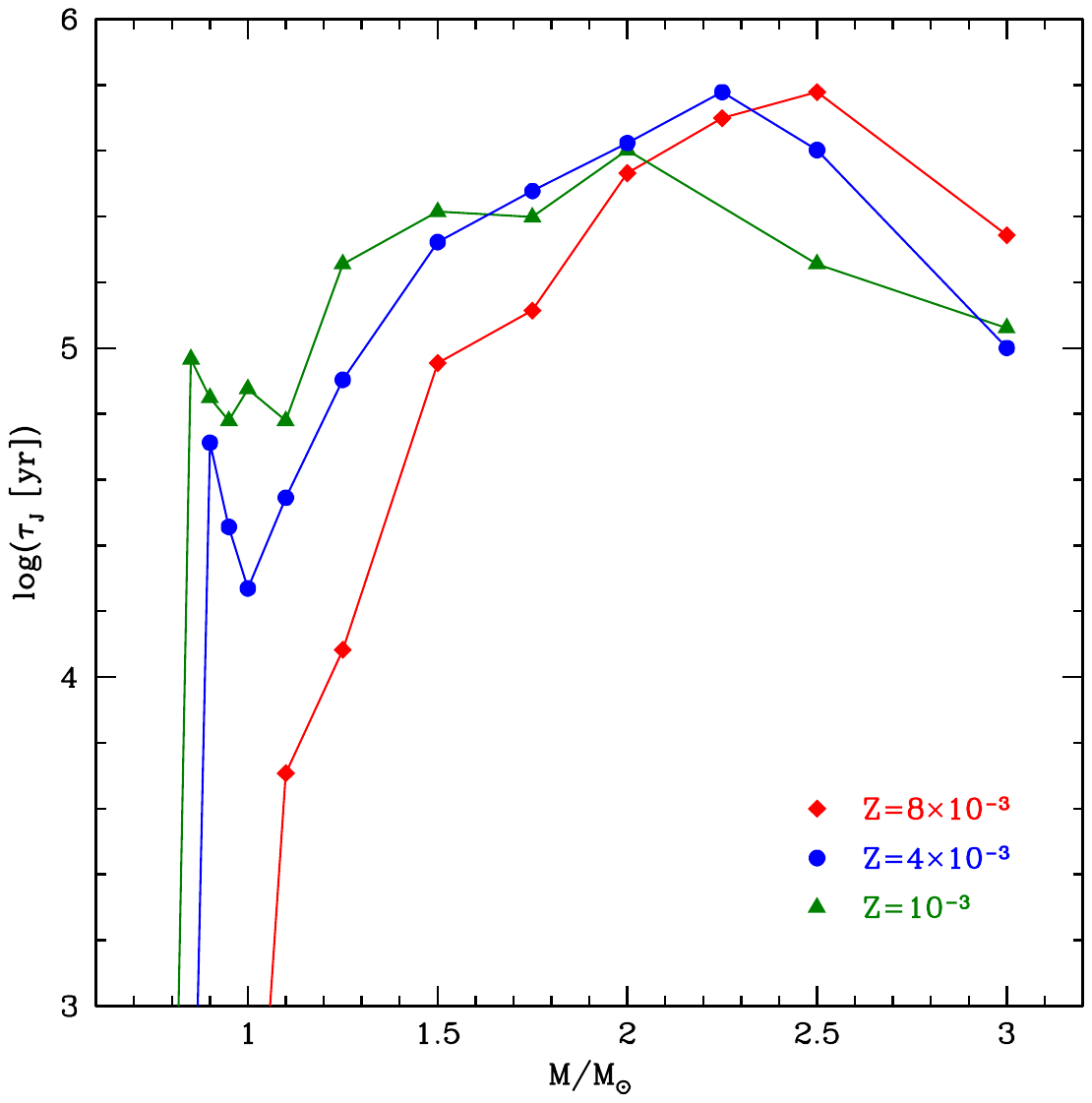}}
\end{minipage}
\begin{minipage}{0.46\textwidth}
\resizebox{1.\hsize}{!}{\includegraphics{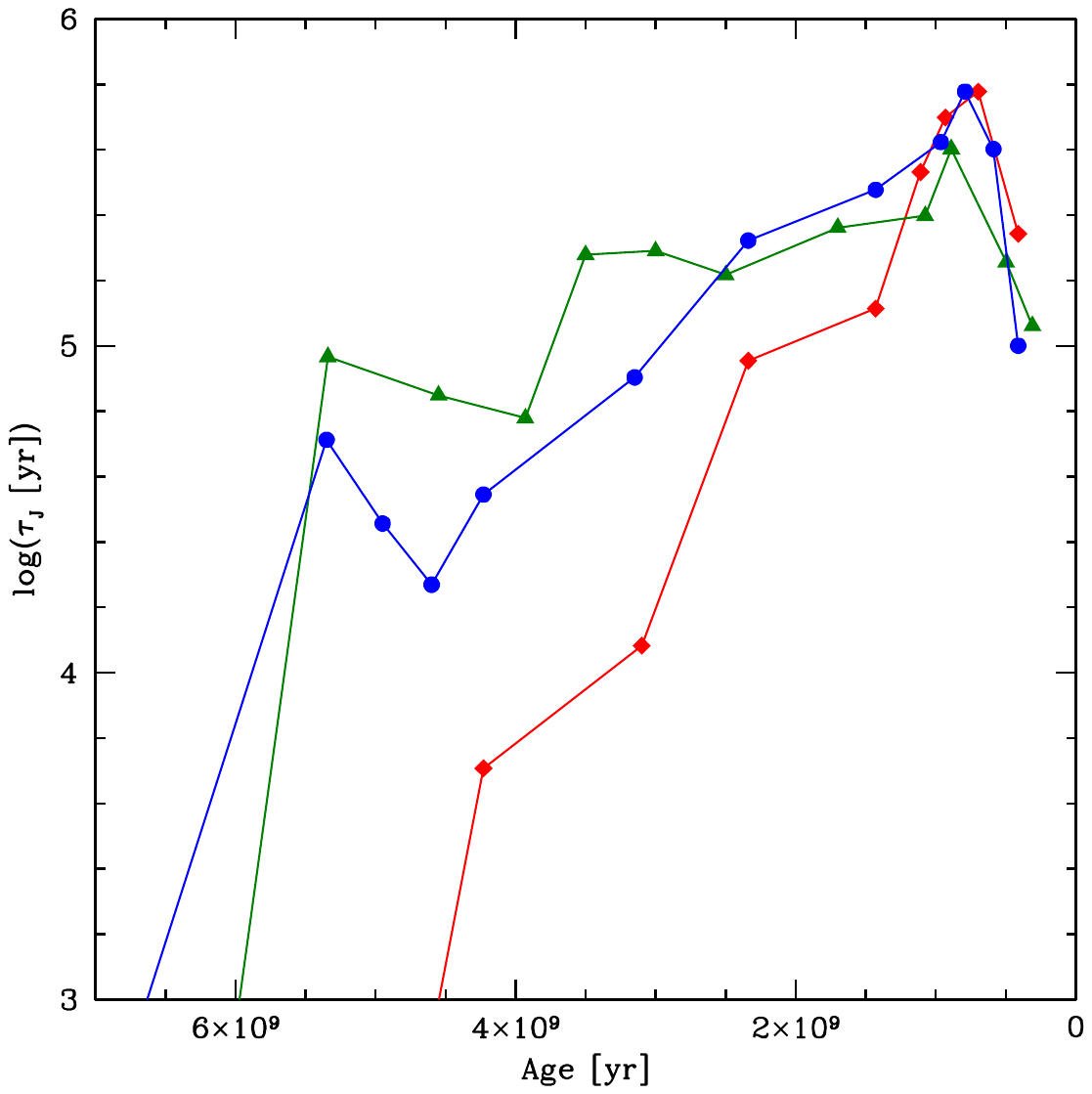}}
\end{minipage}
\vskip-60pt
\caption{Duration of the evolutionary phase spent by model
stars of different metallicity in the box delimiting the
J region of the $\rm (J-K,J)$ plane, according to the
definition by \citet{magnus24}, as a function of the stellar mass
(left panel) and age (right). The masses of $\rm M < 1.5~M_{\odot}$
stars reported in the abscissa of the left panel refer to the 
values at the start of the core helium burning phase.
}
\label{ftempi}
\end{figure*}

Turning to the $\rm 4~M_{\odot}$ model star, whose evolutionary
track is reported in cyan in Fig.~\ref{ftracce}, the results reported
in Fig.~\ref{f40} suggest that massive AGBs evolve into the
$\rm 1.5 < J-K < 2.0$ mag region of the plane during the early
phases following the ignition of HBB, when the depletion of the
surface carbon and the synthesis of nitrogen begins. During this 
period the luminosity increases as a reaction to the start of HBB,
the star expands and cools, the rate of mass loss increases, so that
the circumstellar envelope, for the reasons discussed earlier in this
section, becomes an environment favourable for the formation of
dust, which in the present case is mostly composed of silicates,
with traces of alumina dust \citep{fg06, flavia14a, ventura14}.
This behaviour can be considered as typical of massive AGBs
belonging to the group III) introduced by \citet{ventura22}, 
provided that the metallicity is solar or slightly sub-solar,
which is the case for the intermediate mass populations of
the LMC and SMC.  

The results shown in Fig.~\ref{f40}, when compared
to those reported in Fig.~\ref{f25}, lead to two important
conclusions. First, the time spent by massive AGBs in the
$\rm 1.5 < J-K < 2.0$ mag region of the CMD is significantly 
shorter than for carbon stars
(in the specific cases considered here this time is
$\sim 10^4$ yr and $\sim 6\times 10^4$ yr for the 
$\rm 4~M_{\odot}$ and $\rm 2.5~M_{\odot}$ model stars, 
respectively). An additional important conclusion drawn from
inspection of Fig.~\ref{f25} and Fig.~\ref{f40} is that
the J magnitudes of the stars belonging to the groups II) and
III) are significantly different when they enter the J region.
The inclusion of massive AGBs in the J region of the CMD is 
sensitive to the choice done for the J extension of the 
selected box: for instance, it is clear from inspection of
Fig.~\ref{ftracce} that when the recommended choice proposed
by \citet{magnus24} is adopted, the $\rm 4~M_{\odot}$ model star
considered here would escape detection within the J region.

\subsection{The AGB population in the J region of the CMD}
\label{times}
The results shown in Fig.~\ref{ftracce} indicate that the choice
of the size of the box defining the J region proposed by
\citet{magnus24} is tailored ad hoc to restrict the attention
on carbon stars only: indeed this choice rules out massive AGBs,
which evolve brighter than the lower J magnitude considered
by \citet{magnus24}, and possible contaminations from low-mass, 
O-rich stars, which, if present, would evolve to lower J fluxes 
than the minimum indicated in the aforementioned study.

A deeper inspection of Fig.~\ref{ftracce} suggests that the slope
of the evolutionary tracks within the J region of the CMD changes
with the progenitor's mass: while the J flux of $\rm M \geq 2~M_{\odot}$
stars keeps constant or even increases during the J phase, the lower mass
counterparts are exposed to a more abrupt transition upon becoming
carbon stars, which leads to a fast drop in the J flux, such that
the evolutionary tracks (see the $\rm 1.1~M_{\odot}$, 
$\rm 1.25~M_{\odot}$ and $\rm 1.5~M_{\odot}$ cases in Fig.~\ref{ftracce}) 
cross the J region with a negative slope, pointing the faint, 
right side of the CMD. 

These differences are related to the modality with which the
transition from M-type to C-star occurs in stars of different
mass. $\rm M \geq 2~M_{\odot}$ stars lose little mass during the
initial AGB phases when they evolve as O-rich objects,
because they are characterised by convective envelopes with masses
above $\rm \sim 1~M_{\odot}$, which guarantee a sufficiently large
gravitational attraction in the surface regions, which prevents
intense mass loss. When reaching the C-star stage, their surface
$\rm C/O$ slightly exceeds unity and then grows gradually, because the
carbon dredged-up during the chain of TDU events is efficiently
diluted with the massive envelope, which prevents dramatic
changes in the surface $\rm C/O$ ratio. These objects are
expected to populate the J region during different inter-pulse
phases, and then to evolve off this zone of the CMD only
after a further series of TDU episodes makes $\rm C/O$ to
grow bigger than $\sim 2.5-3$. This is the situation that we
encountered earlier in this section when discussing the
$\rm 2.5~M_{\odot}$ model star, whose evolution was shown in
Fig.~\ref{f25}. The slope of the evolutionary track within the
J regions stems from the balance between the gradual increase in the
luminosity, due to the growth of the core mass, and the rise
in the surface $\rm C/O$, which favours carbon dust production,
triggering a more and more extended depression of the SED in the J 
spectral region. Because in these stars the increase in $\rm C/O$ 
proceeds gradually, the first effects prevails slightly, 
particularly during the phases immediately after the entrance
into the J region, so that the slope of the evolutionary
tracks within the J region is positive, or flat.

The stars descending from progenitors of mass below $\rm 2~M_{\odot}$
evolve differently, because the mass of the envelope is of a few
tenths of solar masses when the transition to C-star occurs.
This is because, when the AGB phase begins, their envelope is less 
massive than that of the higher mass counterparts ,
and this leads to lower surface gravities, which enhances the
mass loss rate experienced by the star during the O-rich phase.
Under these conditions, the transition to the C-star phase
is generally followed by a significant increase in the surface
carbon content, because dilution with the gas stored in the
external envelope is much less efficient than in the higher
mass carbon stars. In these low mass stars the production of
carbon dust takes place very efficiently since they become 
carbon stars: this results into a significant depression of
the J flux, and is the reason for the negative slope of
the evolutionary tracks in the CMD, as clear in Fig.~\ref{ftracce}.

From these results we understand that the modality with which
the stars enter and evolve into the J region changes notably with the
progenitor's mass. The objects of higher mass, in the $\rm 2-3~M_{\odot}$
range, spend in the J region a significant fraction of the time during 
which they evolve as carbon stars, eventually evolving off that region
only after repeated TDU events lifted the surface carbon to 
quantities that favour efficient dust formation and large IR emission.
Conversely, the evolution of the low-mass counterparts within the
J region appears more as a real transition phase, as the effects of
each TDU event on the surface $\rm C/O$, hence on the DPR, are more
dramatic, so that they easily evolve off the region of the CMD
considered.

These arguments regarding the transit of the stars within the J region of the
CMD are confirmed by the results shown in Fig.~\ref{ftempi}, where
the duration ($\rm \tau_J$) of the stay in the J region of the stars of different 
metallicity is shown as a function of the initial mass and the
age (left and right panel, respectively). In case 
of stars with metallicity $\rm Z=8\times 10^{-3}$, whose evolutionary tracks are shown in
Fig.~\ref{ftracce}, we see in the left panel of Fig.~\ref{ftempi}
that $\rm \tau_J$ is positively correlated with the 
mass of the star and changes from $\sim 5\times 10^3$ yr, for 
$\rm M=1.1~M_{\odot}$, to $\sim 6\times 10^5$ yr, for $\rm M=2.5~M_{\odot}$. 
The positive trend of $\rm \tau_J$ vs mass shows a turning point for the
$\rm 3~M_{\odot}$ model star, because the time scale of the AGB evolution 
of the latter is shorter than that of the lower mass counterparts.
The range of masses considered for the $\rm Z=8\times 10^{-3}$ case is limited
to the $\rm 1.1-3~M_{\odot}$ range: this is because the stars that
at the TRGB have $\rm M < 1.1~M_{\odot}$
never enter the J region, while $\rm M > 3~M_{\odot}$ stars evolve
to the red side of the plane, but they are too bright with respect to the
vertical size of the J box. This limitation indicates that only stars formed
between $\sim 300$ Myr and $\sim 4$ Gyr ago populate the J region 
of the CMD, as shown in the right panel of Fig.~\ref{ftempi}.

The $\rm \tau_J$ vs mass trends for the $\rm Z=10^{-3}$ and $\rm Z=4\times 10^{-3}$
cases shown in Fig.~\ref{ftempi} are generally flatter than $\rm Z=8\times 10^{-3}$,
because the C-star condition is reached more easily, owing to the lower content
of oxygen, so that the lower mass threshold required for the stars to
populate the J region is smaller: we see in Fig.~\ref{ftempi} that the minimum
mass evolving into the J region is $\rm 0.8~M_{\odot}$ and $\rm M=0.9~M_{\odot}$,
for $\rm Z=10^{-3}$ and $\rm Z=4\times 10^{-3}$, respectively.

The variation of $\rm \tau_J$ with the stellar age, shown in the
right panel of Fig.~\ref{ftempi}, exhibits a behaviour similar to
the trend $\rm \tau_J$ vs mass, given the tight relationship
between stellar mass and age. It is clear the importance of the
stellar formation that occurred around one Gyr ago
for the numerical consistency of the population of the J region 
of the plane.

\begin{figure*}
\vskip-40pt
\begin{minipage}{0.46\textwidth}
\resizebox{1.\hsize}{!}{\includegraphics{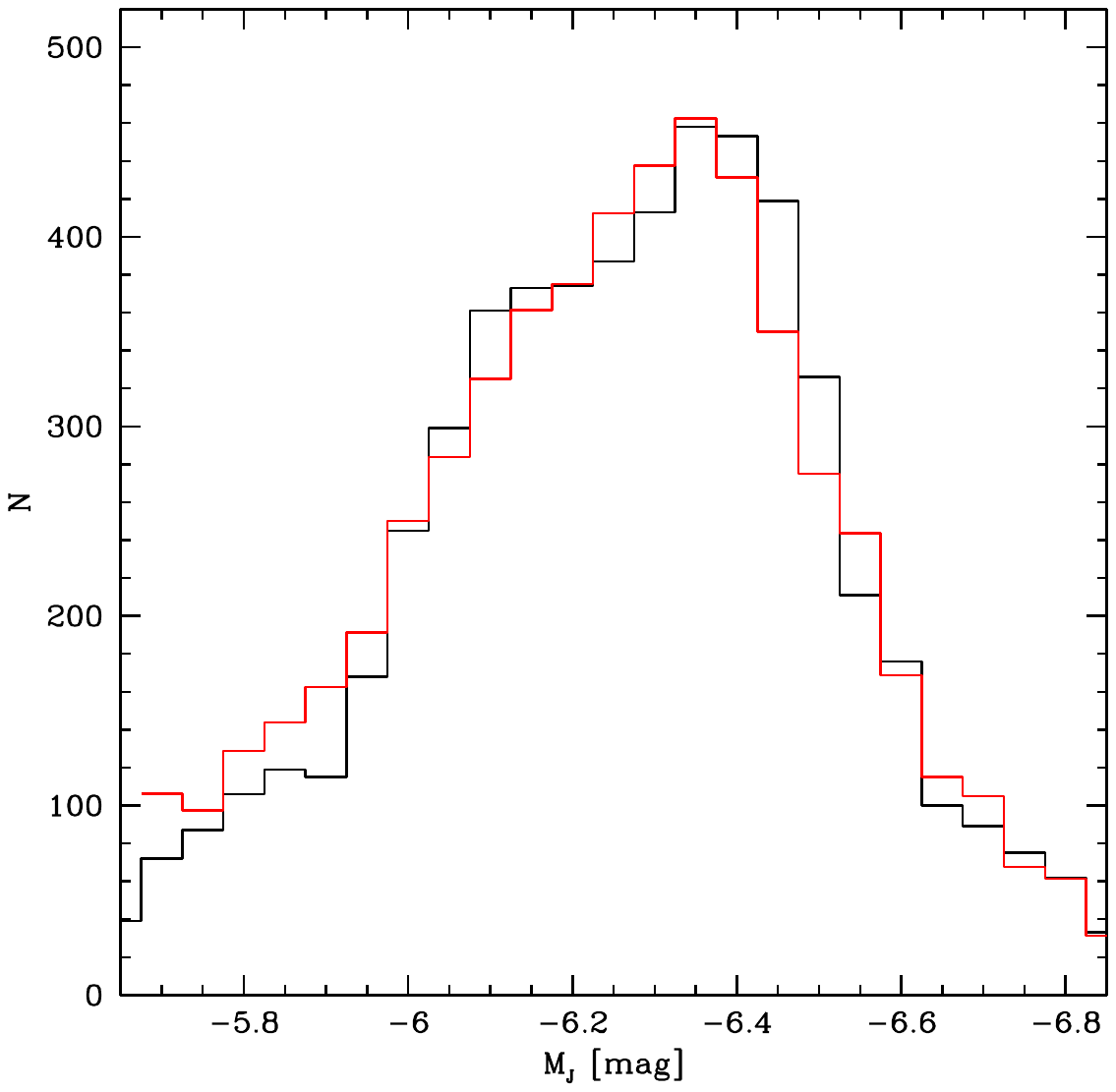}}
\end{minipage}
\begin{minipage}{0.46\textwidth}
\resizebox{1.\hsize}{!}{\includegraphics{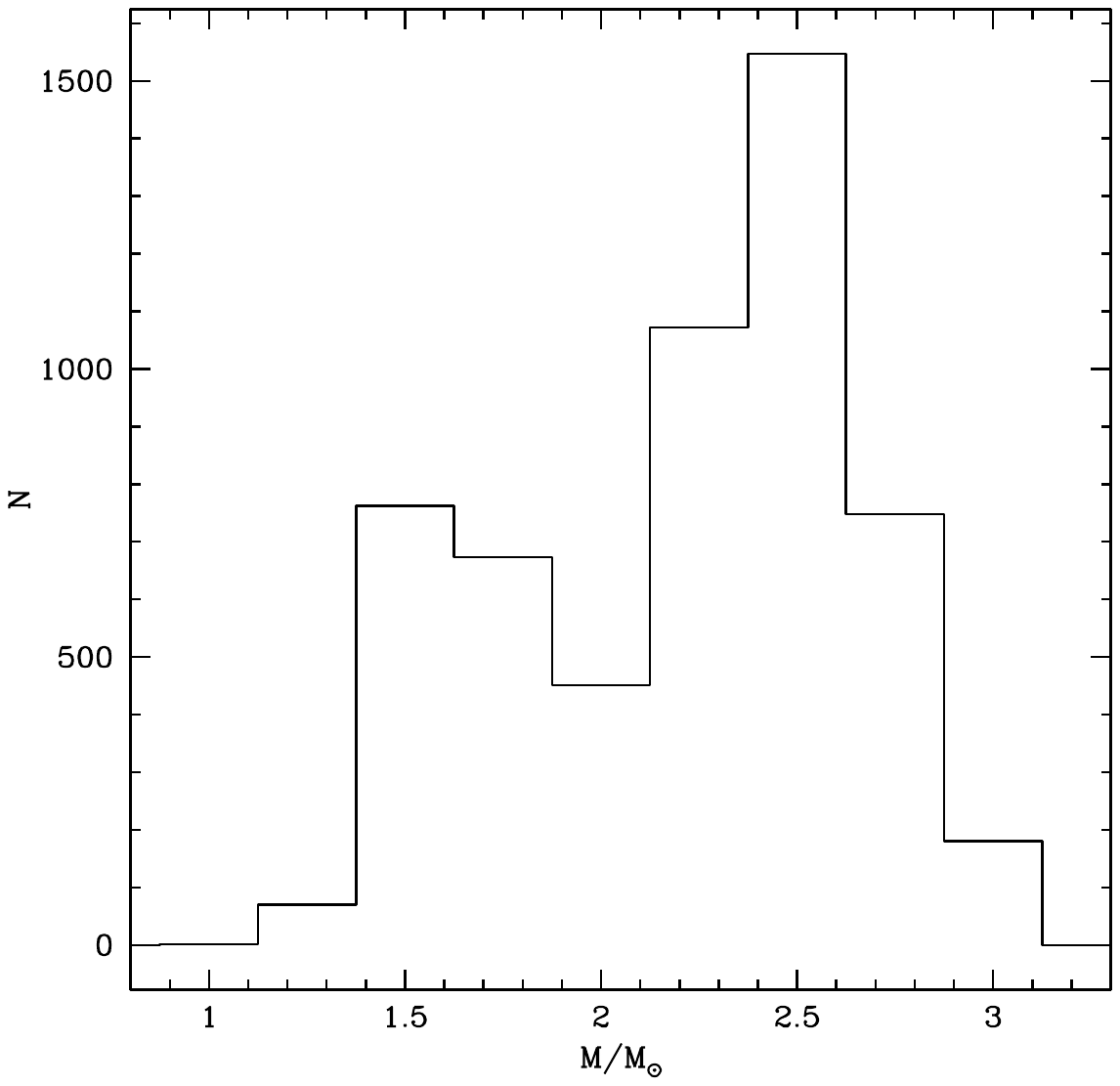}}
\end{minipage}
\vskip-60pt
\caption{{\bf{Left:}} The J luminosity function 
of the LMC AGB stars populating the J region of the $\rm (J-K,J)$ plane, 
obtained by means of population synthesis, is shown in black, and
compared with the JLF by \citet{magnus24} (red line). {\bf Right:} The distribution of the masses of the stars in the J region of the $\rm (J-K,J)$ plane,
which correspond to the JLF shown in the left panel. The masses of $\rm M < 1.5~M_{\odot}$
stars refer to the 
values at the start of the core helium burning phase.
} 
\label{fLMC}
\end{figure*}

\section{The J region stellar population of the LMC}
\label{lmc}
\begin{figure}
\centering
\begin{minipage}{0.5\textwidth}
\resizebox{1.\hsize}{!}{\includegraphics{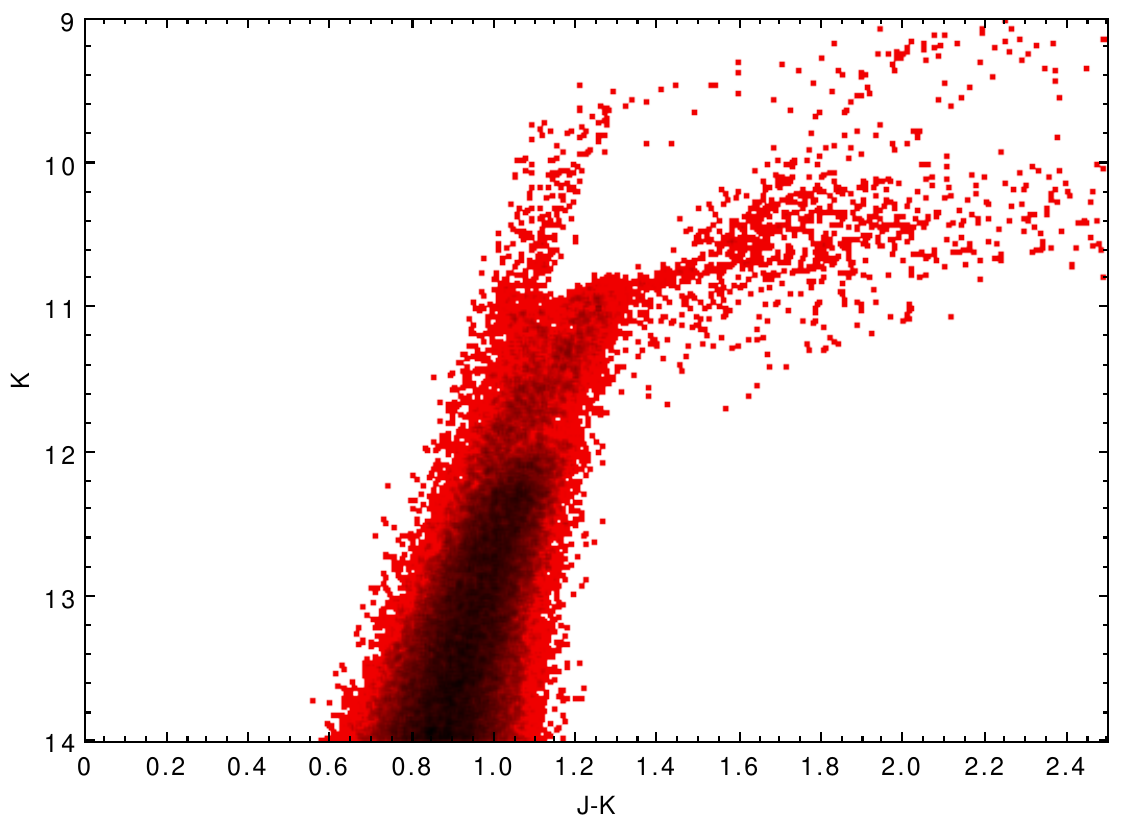}}
\end{minipage}
\caption{Synthetic distribution of LMC stars evolving through the 
AGB phase in the colour-magnitude $\rm (J-K, K)$ plane.} 
\label{fcmd}
\end{figure}

We used the methodology described in section \ref{input} to
build synthetic distributions of LMC stars on the different
observational planes, based on the SFH given
in \citet{mazzi21} and the AMR by \citet{carrera08}.
The results of this population synthesis analysis in the
$\rm (J-K,K)$ CMD is shown in Fig.~\ref{fcmd}. We recognize
in the distribution of the stars the typical features outlined
e.g. by \citet{marigo03}, and particularly: a) the drop in the K-band 
luminosity function at $\rm K \sim 11.9$ mag, where the TRGB
is located; b) the cut-off of the oxygen-rich AGB luminosity function
at $\rm K \sim 10.7$ mag; c) the colour gap between the O and C-rich AGB 
populations, in the $\rm 1.5 < (J-K) < 2.0$ mag region. 
Most of the carbon stars are found in the $\rm (J-K) > 1.2$ mag 
region, although a few C-rich objects are also expected to populate
the AGB branch above the TRGB: these are the progeny of 
$\rm 2-3~M_{\odot}$ stars, which during the phases immediately
following the start of the C-star phase evolve into the blue region
of the CMD (the interested reader can found an exhaustive discussion
on this argument in Gavetti et al. in prep.).

For what concerns specifically the J region, it is clear from
the arguments presented in the the previous section
that the numerical consistency and the distribution of the stars
within this region of the CMD depends on the number
of stars descending from $\rm \sim 1-3~M_{\odot}$ progenitors
nowadays evolving through the AGB phase. Therefore, the SFH
of the galaxy in the epochs between 300 Myr and 4
Gyr ago is the key factor to be considered, along with the 
AMR, particularly the metallicity of the interstellar medium
in those times. These restrictions are partly sensitive to the
criterion adopted to select the J region and hold strictly
for that proposed by \citet{magnus24}. If the J region considered
was extended to brighter J fluxes than the limits proposed by
\citet{magnus24}, stars more massive than $\rm 3~M_{\odot}$ would 
need to be considered, thus even the recent star formation, occurred 
$\sim 100-200$ Myr ago, should be examined. No significant changes
are expected if the chosen box for the J region is extended to fainter
magnitudes, since the present choice is such that, as shown in
Fig.~\ref{ftracce}, even the evolutionary tracks of 
the lowest luminosity carbon stars, i.e. those descending from old, 
low-mass progenitors, cross the J region of the 
$\rm (J-K,J)$ plane; the only exception would be
found in the galaxies characterised by the presence of an old, 
metal-rich stellar population, as under those conditions 
low-mass, oxygen-rich stars during the final AGB phases produce 
amounts of silicate sufficiently large to deviate the evolutionary
tracks towards the faint side of the $\rm 1.5 < J-K < 2.0$ mag strip
of the CMD. However, this is not the case neither for the LMC, 
nor for the SMC, which will be discussed in next section.

The J luminosity function (JLF) obtained for the LMC by means of the
population synthesis method is reported 
as a black line in the left panel of 
Fig.~\ref{fLMC}, while the distribution of the (initial) masses 
of the stars populating the J region is shown in the right
panel of the figure.

Inspection of Fig.~\ref{fLMC} shows a satisfactory agreement between
the present findings and the results shown in the
Fig.~F3 in \citet{magnus24}, particularly for what attains the
peak of the distribution at $\rm M_J=-6.35$ mag and the general shape 
of the JLF, which drops to half the peak value at $\rm M_J=-6.55$ mag 
and $\rm M_J=-6$ mag on the bright and faint sides of the distribution, 
respectively. Full consistency is also found for the average
J magnitude, which is around $\rm M_J=-6.25$ mag in both
cases.

The mass distribution shown in the right panel of Fig.~\ref{fLMC}
shows up two peaks, centered at $\rm 1.7~M_{\odot}$ and 
$\rm 2.5~M_{\odot}$, related to stars formed during the peaks in the SFH 
of the LMC occurred 1.5 Gyr and 800 Myr ago \citep{mazzi21}, respectively. 
The older peak in the SFH is higher than the younger one, so that 
$\rm 1.5 \leq M <1.8~M_{\odot}$ stars currently evolving along the AGB 
outnumber by almost a factor two the $\rm 2.2 - 2.7~M_{\odot}$ counterparts. 
However, this is not the case for the J region, because, as shown in 
Fig.~\ref{ftempi}, the latter stars 
spend a significant fraction  of the AGB lifetime within that area of 
the CMD, thus they provide the dominant contribution to the overall 
population of the J region: the peak in the JLF at $\rm M_J=-6.35$ mag,
and also the two nearest bins in the distribution, are almost entirely
due to the presence of $\rm \sim 2.5~M_{\odot}$ stars.
The asymmetry of the JLF around the peak
value is due to the excess of $\rm 1.5-2~M_{\odot}$ stars, which provide
the dominant contribution to the faint side of the distribution, 
with respect to $\rm M >2.5~M_{\odot}$ stars, which contribute to the
bright side of the JLF.

The present analysis indicates that about $2/3$ 
of the stars in the J region of the LMC descend from $\rm \sim 2.5~M_{\odot}$
progenitors, formed between 700 Myr and 1 Gyr ago. These stars
are characterised by surface $\rm C/O$ ratios in the $1.5-3$ range 
and are producing carbon dust at rates 
$\rm \sim 2-3 \times 10^{-9}~M_{\odot}/$yr, distributed between solid
carbon dust, which accounts for a fraction of the total dust formed
between $50 \%$ and $75 \%$, and silicon carbide (SiC), whose contribution
is between $25 \%$ and $50 \%$. Solid iron and other dust species
are formed in minor quantities and do not affect the shape of the SED.
The luminosities of these
objects are in the $\rm 7000-9000~L_{\odot}$ range, which correspond
to J magnitudes $\rm -6.6 < M_J < -6.2$: as shown in Fig.~\ref{ftracce},
they populate the upper part of the J region. 

\citet{flavia14b, flavia15a} suggested that the sources of the
LMC with the largest IR emission, which populate the reddest regions of the
observational planes built with the IRAC Spitzer filters, are the progeny
of $\rm 2.5-3~M_{\odot}$ stars, and are currently providing the dominant
contribution to the overall DPR from the entire galaxy. From the present
investigation we deduce that the stars that provide the most relevant
contribution to the population of the J region of the LMC are the
immediate progenitors of the reddest objects investigated by
\citet{flavia14b, flavia15a}.

A significant fraction of the stars in the J region, around $30\%$, are
the progeny of $\rm 1.5-1.8~M_{\odot}$ stars formed during the epochs
from 1.5 to 2.5 Gyr ago. The surface $\rm C/O$ of these objects is
between 1.5 and 2, the DPRs are in the 
$\rm 2-3 \times 10^{-9}~M_{\odot}/$yr range, with relative contributions
from solid carbon and SiC not significantly different from those
given above for the higher mass stars. These stars, with
luminosities between $\rm 5\times 10^3~L_{\odot}$ and 
$\rm 7\times 10^3~L_{\odot}$,  provide
the most relevant contribution to the JLF at $\rm -6.3 < M_J < -6$
and are mostly located in the fainter side of the J box.

The AGB population of the LMC within the J region also encompasses
$\sim 10\%$ of stars formed around 300 Myr ago from $\rm \sim 3~M_{\odot}$
progenitors, located in the bright side of the J region (see 
Fig.~\ref{ftracce}), where they evolve
at luminosities slightly above $\rm 10^4~L_{\odot}$. Among the
sources in the J region these bright stars are those producing dust at
the largest rates of $\rm \sim 5\times 10^{-9}~M_{\odot}/$yr.

Towards the low luminosity tail of the JLF shown in the left panel of 
Fig.~\ref{fLMC}, particularly in the $\rm M_J>-6$ mag region, we find
a group of stars descending from progenitors of mass slightly
super solar, which account for $\sim 2\%$ of the whole population in
the J region of the plane. The reason why only a few of these objects
are found nowadays in the J region is related to the rapidity with which
the evolutionary tracks cross this part of the plane, for the arguments
discussed in section \ref{times}. These are the oldest among the LMC stars
considered here, and formed 3-4 Gyr ago. 

\subsection{The role of the RGB mass loss}
According to \citet{mazzi21}, some star formation took place in the
LMC in epochs older than 4 Gyr, when stars of mass below 
$\rm \sim 1.2~M_{\odot}$, now evolving through the AGB, formed.
In contrast to their younger and more massive counterparts, the 
AGB evolution of these stars - specifically their dust-production 
efficiency and the resulting displacement of their evolutionary 
tracks across the observational planes considered - is strongly 
influenced by $\rm \delta M_{RGB}$, the mass lost during the RGB 
phase prior to helium-flash ignition.
In a recent
work on the evolved stellar population of M31, \citet{cla} outlined
the relevant role of $\rm \delta M_{RGB}$ in shaping the luminosity
function of the stars of the galaxy populating the blue side of the
CMDs obtained with some HST filters, accounting for $\sim 90\%$
of the entire sample. One of the conclusions drawn by \citet{cla} is
that for stars of solar or sub-solar mass, 
RGB mass losses $\rm \delta M_{RGB} \sim 0.2-0.3~M_{\odot}$ are required to 
reproduce the observed luminosity function, the detailed value 
of $\rm \delta M_{RGB}$ increasing with metallicity. 

As regards the analysis in the present investigation, 
the choice of $\rm \delta M_{RGB}$ is potentially affecting 
the statistics of the AGB population within the J region, and
particularly the last point discussed in the previous 
sub-section, that the oldest stars now sampled in the J region 
of the LMC formed $\sim 4$ Gyr ago. This is because
the path towards the formation of carbon stars of low-mass
objects depends on the mass of the envelope when the AGB phase 
begins: in the metal-poor domain, which must be considered
if the oldest epochs are taken into account, all the stars that
reach the AGB with mass $\rm M \geq 0.8~M_{\odot}$ become
C-stars \citep{devika23}, so that their evolutionary tracks evolve to the red,
and thus enter the J region of the CMD. When mass loss
during the RGB is considered, the threshold initial mass of the 
star required to become carbon star is higher than the value given
above, which restricts within the J region the mass range of the stars potentially 
able to evolve.

As stated in section \ref{input}, in the analysis developed
here we considered the results by \citet{cla} and assumed 
$\rm \delta M_{RGB} = 0.2~M_{\odot}$ for all the stars
of initial mass $\rm M < 1.5~M_{\odot}$. This choice rules
out the possibility that these objects, which on the basis of
the SFH adopted we estimate to account for
$\sim 30\%$  of the AGB population of the LMC, form carbon dust
and evolve to the red side of the CMD, and leads to the conclusion 
that only the stars younger than 4 Gyr ago must be considered for the
statistical analysis of the J region. 

\begin{figure}
\vskip-40pt
\centering
\begin{minipage}{0.5\textwidth}
\resizebox{1.\hsize}{!}{\includegraphics{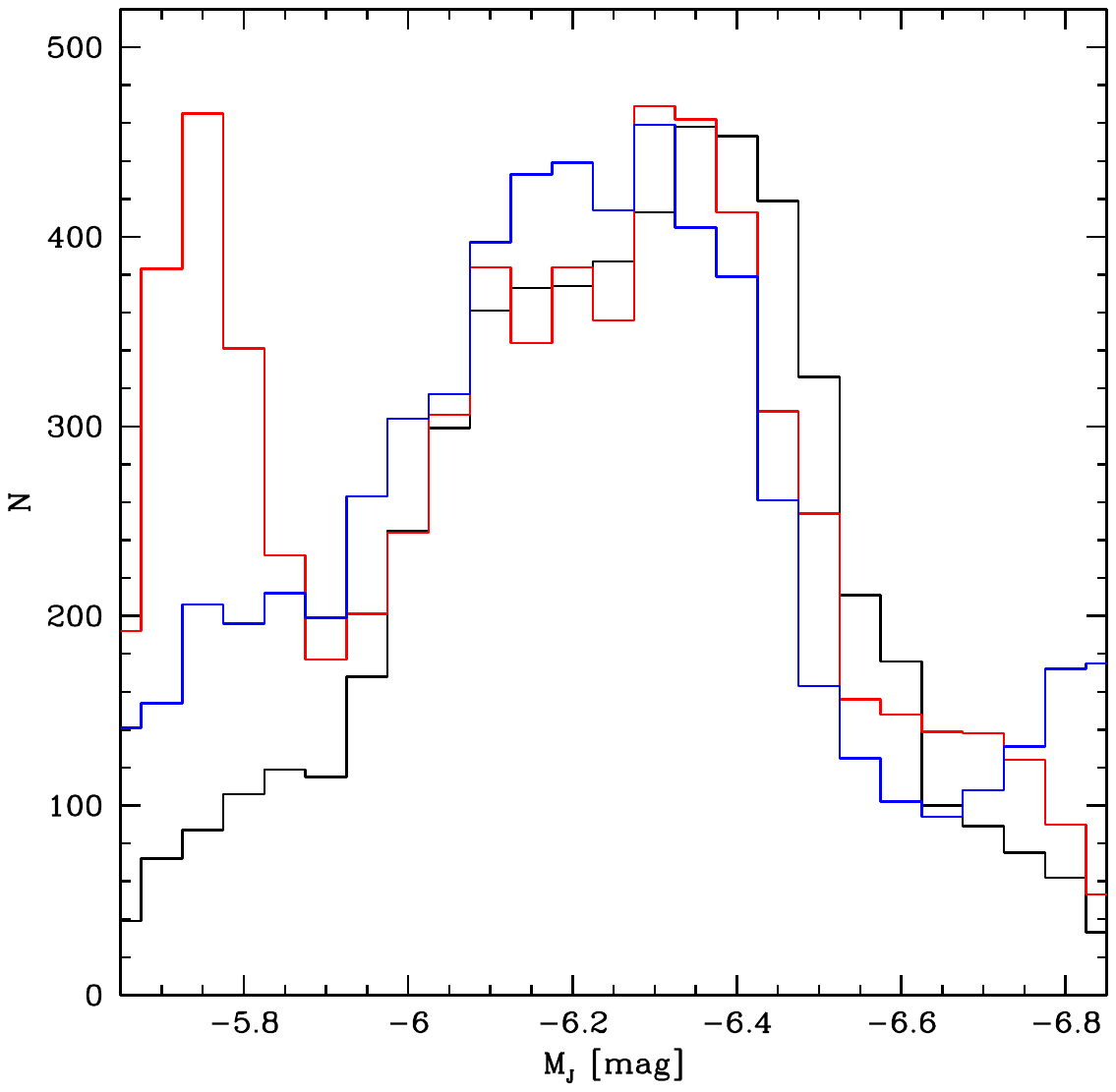}}
\end{minipage}
\vskip-70pt
\caption{The J luminosity function of the stars in the J region
of the LMC, derived on the bases of the assumptions discussed in 
section \ref{input} (black line), is compared with the results obtained
when the metallicity of the stars is artificially kept to
$\rm Z=10^{-3}$ until presently (blue line) and when RGB mass loss
is neglected (red).
} 
\label{fconf}
\end{figure}

To understand the potential impact of $\rm \delta M_{RGB}$ on the
synthetic JLF of the LMC, we ran numerical simulations based on
the assumption that no mass loss occurred during the RGB evolution. 
The results, shown in Fig.~\ref{fconf}, indicate that in this case the JLF 
would exhibit a low-luminosity peak, whose height is comparable to,
though lower than, the main peak discussed above. While the 
$\rm \delta M_{RGB} = 0$ assumption is not realistic, we 
deduce that a minimum RGB mass loss is required to reproduce
the results by \citet{magnus24}, and that these findings are
consistent with those by \citet{cla}.

\subsection{The role of the metallicity enrichment}
The study by \citet{carrera08} shows that LMC
stars have a mean metallicity of $\rm [Fe/H]=-0.5$ and 
that the stars younger than 3 Gyr formed with $\rm [Fe/H]>-0.5$. 
Based on the discussion presented earlier in this section, we know that these
slighty sub-solar chemistries characterise the stars within the
J region of the CMD, of interest here.

We checked the role played by the chosen AMR, by repeating the 
population synthesis simulation, assuming that the metallicity of 
the stars is left unchanged with time and is kept at $\rm Z=10^{-3}$ until 
the present epoch. This experiment is not only intended to test 
the uncertainties related to the assumed AMR, but also, in a future
perspective, to understand which situation must be expected when
metal-poor galaxies will be considered.

The results obtained are shown in Fig.~\ref{fconf}, where the derived JLF 
(blue line) is compared with the JLF discussed in the first part of this 
section (reported in black), which reproduces the results by \citet{magnus24}.
While we note some similarities, particularly in the location of the
peak at $\rm M_J \sim -6.3$ mag and in the morphology of the distribution on
the bright side, we also detect some differences that concern the
faint side, where the ''metal-poor'' JLF is significantly flatter than
that based on the AMR by \citet{carrera08}. This result, apparently
in contrast with the general idea that metal-poor stars evolve at
higher luminosities than the lower metallicity counterparts of the
same mass, is explained by the fact that metal-poor stars become carbon stars
more easily \citep{devika23}, owing to the lower initial oxygen
content. As regards the case discussed here, we find that when
the low metallicities are adopted, the stars of mass $\rm \sim 1.5~M_{\odot}$
reach the C-star stage in a less advanced AGB phase than their higher
metallicity counterparts of the same mass, thus they spend a longer time 
within the J box, so their behaviour is more similar to that of the 
higher mass counterparts discussed in section \ref{times}. These stars
are responsible for the bump in the JLF in the
$-6.2 < \rm M_J < -6.1$ range, seen in Fig.~\ref{fconf}. 
A further difference introduced by the assumption that the
stellar population of the LMC is metal-poor is found in the
low luminosity tail of the JLF, at $\rm M_J \sim -5.8$ mag. This is
due to the presence of low mass ($\rm 1-1.3~M_{\odot}$) stars,
which in the previous simulation barely entered the J box,
while in the present case, due to the easier modality with 
which they become C-stars, provide a significant contribution 
to the population on the faint side of the box identifying the
J region of the CMD.

\section{A comparative analysis of the SMC versus the LMC}
\label{smc}
The same approach based on population synthesis used for the
study of the LMC was applied to investigate the J population of 
the SMC and compare the synthetic JLF with that given in
\citet{magnus24}. As discussed in section \ref{input}, in this case 
we adopted the SFH and the AMR given in \citet{rubele18}.

The star formation process in the two galaxies proceeded with
different modalities. First, the metal enrichment in the LMC
was more efficient than in the SMC, which reflects in a lower
average $\rm [Fe/H]$ for the latter galaxy, which affects in
particular the stellar population formed over the last 2 Gyr. 
Furthermore, while the SFH of the LMC is characterised by the
two peaks mentioned earlier in this section, which occurred 
800 Myr and 1.5 Gyr ago, in the SMC there was a single period of 
intense activity between 4 Gyr and 6 Gyr ago,
when the rate of star formation was about
a factor of 2 higher than that experienced 
between 300 Myr and 2 Gyr ago \citep{rubele18}. 
Because of these dissimilarities between the histories of the two galaxies,
the stellar populations nowadays evolving along the AGB
are significantly different: the AGB population of the SMC is 
dominated by stars descending from $\rm 1-1.5~M_{\odot}$ progenitors,
formed during the times running from 6 Gyr ago to 2 Gyr ago, while
for the LMC we underlined that the majority of the AGB stars 
descend from $\rm 1.5-1.8~M_{\odot}$ stars formed 1-2 Gyr ago.
It is important to understand if and 
how these differences affect the stellar population of the two galaxies 
in the J region of the CMD.

\begin{figure}
\vskip-40pt
\centering
\begin{minipage}{0.5\textwidth}
\resizebox{1.\hsize}{!}{\includegraphics{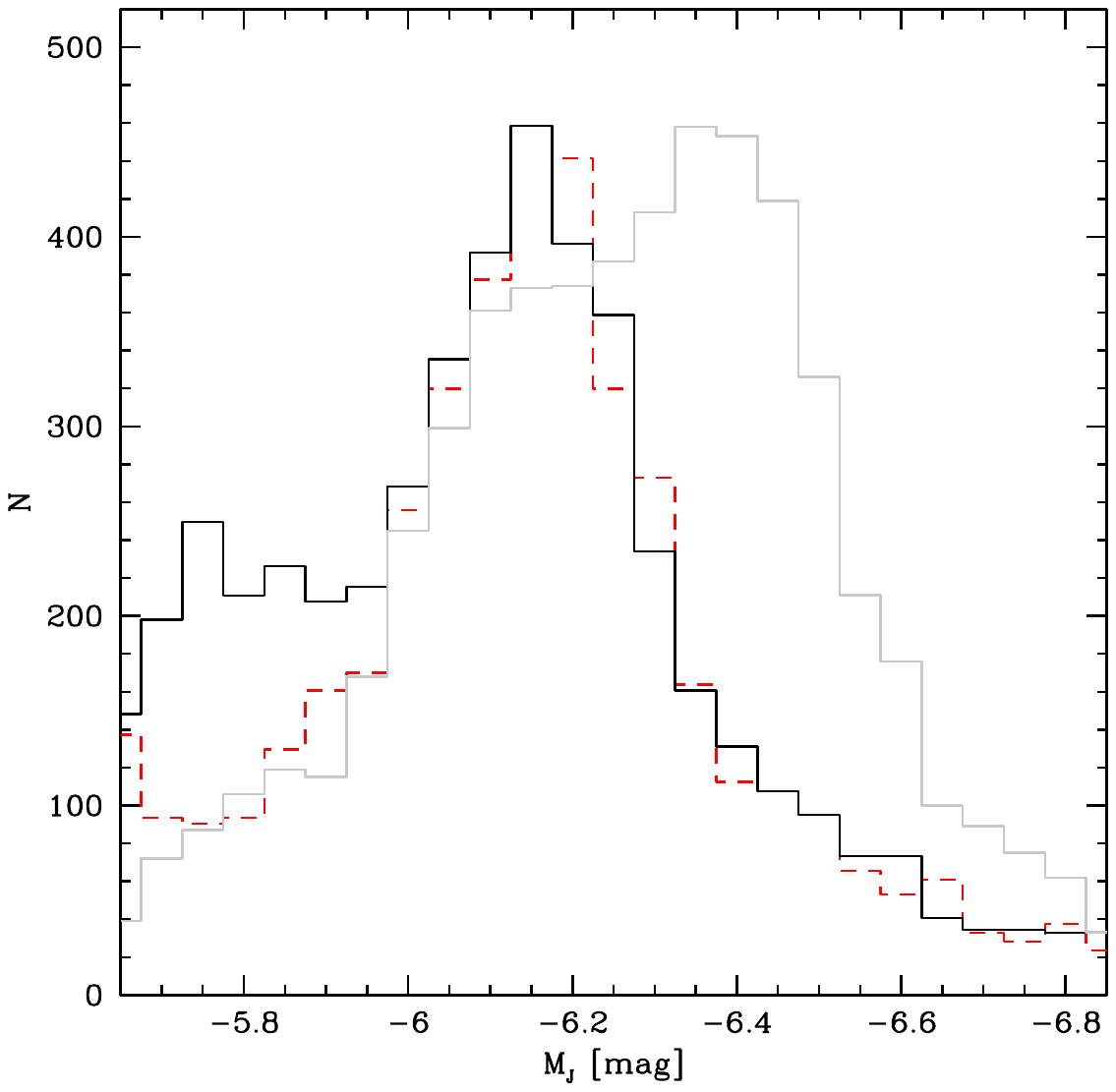}}
\end{minipage}
\vskip-70pt
\caption{The comparison between the J luminosity functions 
obtained for the LMC (grey line) and the SMC (black). The red,
dashed line indicates the results obtained for the SMC on the basis of the
assumption that $\rm 1-1.3~M_{\odot}$ metal-poor stars experience
a $\rm 0.3~M_{\odot}$ mass loss during the RGB evolution.} 
\label{fsmc}
\end{figure}

Fig.~\ref{fsmc} shows the comparison between the JLF obtained
for the SMC, indicated with the solid black line, and the corresponding
distribution obtained for the LMC, discussed earlier in this
section, reported in grey. The JLF of the SMC was artificially 
scaled to ease the comparison between the two distributions.
The first clear difference between the JLF
of the two galaxies is
the location of the peak, which is almost 0.2 mag fainter in the
SMC than in the LMC. This is not surprising, considering that 
the peak of the JLF at $\rm M_J = -6.35$ mag of the LMC was 
determined by the presence of a large number of stars descending 
from $\rm \sim 2.5~M_{\odot}$ progenitors; conversely, the peak in 
the JLF of the SMC, located at $\rm M_J = -6.15$ mag, is
due to the progeny of $\rm 1.2-1.8~M_{\odot}$ stars, which 
are generally fainter than the $\rm 2.5~M_{\odot}$ stars
populating the J region of the LMC (see Fig.~\ref{ftracce}).

Not only the peak, but the whole JLF of the SMC is shifted towards 
smaller J fluxes in comparison to the LMC, owing to the smaller mass 
of the progenitors populating the AGB of the SMC stars with respect
to the LMC, and more specifically the J region of the CMD: while in 
the LMC, as shown in the right panel of Fig.~\ref{fLMC}, we find that 
the sources populating the J region are mainly 
$\rm 2~M_{\odot} < M < 3~M_{\odot}$ stars, the mass distribution of the 
SMC is more homogeneous, with a majority presence of 
$\rm 1~M_{\odot} < M < 1.5~M_{\odot}$ objects.
The presence of such a significant fraction of low-mass stars
is also the reason for the presence of the low luminosity tail 
in the $\rm -5.9 < M_J < -5.7$ mag range of the JLF, which can be seen
in Fig.~\ref{fsmc}. 

The comparison between the JLF obtained by means of population
synthesis and the results from \citet{magnus24} are less
straightforward and indicative than for the LMC, 
given the low number of sources on which the statistical analysis by
\citet{magnus24} is based. The JLF reported in Fig.~\ref{fsmc}
is similar to the one shown in Fig.~F3 by \citet{magnus24} for what
concerns the location of the peak and the general shape of the JLF on 
the bright side of the distribution. Differences are found
on the faint side of the JLF, mainly because of the aforementioned
low luminosity tail, which is not so evident in the results by 
\citet{magnus24}. This tail is also the reason for the difference
in the average J magnitude, which is found to be $\rm M_J=-6.1$ mag in
the present analysis, whereas \citet{magnus24} find $\rm M_J=-6.18$ mag.

The tail on the faint side of the synthetic distribution is due to 
the presence of a population of old, metal-poor 
stars in the J region of the SMC, which descend from progenitors of 
mass slightly above solar. The possibility that these stars enter the
J region is tightly linked to the mass lost during the RGB, $\rm \delta M_{RGB}$: 
significant RGB mass loss reduces the mass with which the stars start the AGB, 
so that they undergo an evolutionary path similar to that of the stars
belonging to the group I) discussed in section \ref{gruppi}, with no
chance of populating the J region of the plane.
The results discussed above, reported with the black, solid line in Fig.~\ref{fsmc}, were
obtained with a mass loss $\rm \delta M_{RGB}=0.2~M_{\odot}$
for stars of mass below $\rm M=1.5~M_{\odot}$. If we assume
$\rm \delta M_{RGB}=0.3~M_{\odot}$ for all $\rm M<1.5~M_{\odot}$ stars,
we obtain the JLF indicated with the red, dashed line in Fig.~\ref{fsmc}.
This is in a much better agreement with the results by \citet{magnus24},
as also confirmed by the derived average J flux, which is $\rm M_J=-6.16$ mag.
The choice of $\rm \delta M_{RGB}$ does not significantly affect the location of the 
peak of the JLF, because the bulk of the population of the J region of
the SMC is made up of the progeny of $\rm 1.2-1.8~M_{\odot}$ stars, which are
touched only marginally by the assumption of $\rm \delta M_{RGB}$, except for
the smaller masses in the mass interval. In a more general context, the
results by \citet{freedman20} and \citet{Lee24a, Lee24b} seem to rule out
the presence of low luminosity tails in the JLF of all the galaxies investigated: 
this suggests that significant mass loss occurred during the RGB phase of low mass 
stars, in agreement with the conclusions given above.

The issue of the mass loss experienced by the stars during the RGB phase is much 
more relevant for the SMC than for the LMC, because the SMC hosts a much higher
fraction of low-mass stars, whose AGB evolution is extremely sensitive to
the assumed $\rm \delta M_{RGB}$, which inevitably reflects on the statistics
of the masses and luminosities of the population of the J region. The smaller
metallicity of SMC stars is a further motivation to take particular care of the
mass loss process during the RGB, because metal-poor stars reach the C-star
stage more easily, thus it is more likely that they eventually enter the
J region.

The present analysis on the properties of the JLF, although limited
to only two galaxies, is quite exhaustive, because the LMC and the SMC 
likely represent the two extreme cases that can occur, at least as far 
as the peaks of the distribution are concerned. It is hard to think
that there could be galaxies whose JLF peaks brighter
than the LMC, i.e. $-6.35$ mag, which we have seen to correspond to the average
J magnitude of $\rm \sim 2.5~M_{\odot}$ stars, when they evolve within the
J region. A brighter peak would demand a dominant population of stars
of $\rm 3~M_{\odot}$ or more, which is hard to believe, even if intense
star formation occurred during the epochs younger than 300 Myr: indeed 
the AGB evolutionary time scales drop significantly in the 
$\rm M > 2.5~M_{\odot}$ mass domain, and the $\rm M_J$ vs mass
relationship becomes steeper and steeper in that mass range.
This conclusion holds even more if the choice recommended by 
\citet{magnus24} to identify the J region is adopted, since
$\rm M > 3~M_{\odot}$ stars are not expected to populate the
J region, as shown in Fig.~\ref{ftracce}.

The SMC represents the opposite extreme case, in that we 
consider rather unlikely the presence of galaxies whose 
peak of the JLF is fainter than that of the SMC, i.e. $-6.15$ mag. 
Earlier in this section we discussed
that the reason for this rather faint peak, at least 
when compared to the LMC, is the older and lower mass population 
of the SMC than the LMC, mostly formed around 5 Gyr ago. 
Even an older burst in the star formation activity 
would barely change the location of the peak of the JLF,  because
that would favour the formation of low-mass stars that hardly
enter the J region, as they mostly evolve as the stars belonging
to the group I) discussed in section \ref{gruppi}. 

Assessing whether the peak J ($\rm M_J^{peak}$) or rather the 
average J magnitude ($\rm M_J^{av}$)
of the stars located in the J region of the plane should be used as 
distance indicator is not an easy task, at least on the basis of the
present results, which are restricted to two galaxies only.
Use of the $\rm M_J^{peak}$ has the advantage that the attention can
be focused on a narrow range of masses (hence formation epochs),
with a strong bias towards the stars formed around 1 Gyr ago. 
With the exception of those cases where a burst of star formation
occurred 4 Gyr ago (which is the case for the SMC), we expect
to find $\rm M_J^{peak} \sim -6.3$ mag. On the other hand,
the average J magnitude is less sensitive to the details of
the SFH between 800 Myr and 4 Gyr ago, as it is affected by
the J fluxes of a wider range of stellar masses, whose evolutionary tracks 
enter the J region. The differences arising from variation in
the SFH are softened in this case, as confirmed by the
similarity between the values found for the LMC and SMC,
which are much closer ($\rm \delta M_J^{av} \sim 0.07$ mag) 
than the corresponding 
peak magnitudes ($\rm \delta M_J^{peak} \sim 0.2$ mag).
The shortcoming of the use of $\rm M_J^{av}$ is
that the results are affected by the 
uncertainties connected to the modelling of the low mass
stars, whose presence has practically no influence on the
determination of the peak value. By adopting the average
J magnitude one is forced to come across the issues of the
treatment of mass loss along the RGB, a physical mechanism
highly uncertain, but with a generally strong impact on the 
statistics for the AGB population of galaxies \citep{cla}.

\section{Conclusions}
\label{concl}
We use a population synthesis approach to investigate the
distribution of MCs AGB stars that
occupy the so-called J region of the $\rm (J-K,J)$ colour-magnitude
 diagram. This analysis is stimulated by recent studies suggesting that the J-band luminosity function of stars
in that region of the plane may serve as distance indicators
of galaxies.

Based on the morphology of the evolutionary tracks of stars
of various mass and chemical composition on the $\rm (J-K,J)$ plane,
we find that the J population of the galaxies descend from
$\rm 1-3~M_{\odot}$ stars, between 300 Myr and 5 Gyr old,
caught in the evolutionary phases between the stage when
they become carbon stars and the phase when the surface
$\rm C/O \sim 3$, the latter condition favouring a significant shift
of the SED to the IR spectral region, so that the stars
evolve off the J region. 

The timing of the transit of stars through the J region is 
extremely sensitive to the mass of the star: while $\rm M > 2~M_{\odot}$
stars remain in that zone of the plane for a series of 5-6
inter-pulse phases, for a total duration slightly shorter than
1 Myr, the lower-mass counterparts evolve faster, and rapidly evolve 
to the red side of the plane. In light of this, we expect that
in several galaxies the J luminosity function of the stars
populating the J region peaks at $\rm M_J \sim -6.3$ mag, which is the
J magnitude of $\rm 2.5~M_{\odot}$ stars during the evolution in
that part of the CMD.

The results concerning the LMC are satisfactory and confirm the conclusion
given above. The simulation obtained by applying the population synthesis
method is fully consistent with the observational scenario, as
both the peak of the distribution, found at $\rm M_J \sim -6.35$ mag, and
the width of the J luminosity function on both sides of the peak, 
are nicely reproduced. Mass loss during the RGB evolution of the order of $
\rm 0.2~M_{\odot}$, in full agreement with previous investigations, must be 
invoked to prevent the appearance of an unobserved extended tail on the faint 
side of the J luminosity function.

The J luminosity function of the SMC is shifted to higher $\rm M_J$'s than 
the LMC, the main
peak being located at $\rm M_J \sim -6.15$ mag. We interpret this difference
as due to the intense star formation that occurred in the SMC around
5 Gyr ago, when stars of mass slightly above solar formed and are now
evolving along the AGB. Under these specific conditions these stars
outnumber the $\rm 2-3~M_{\odot}$ counterparts, so the peak of the
distribution of the J magnitudes occurs at fainter J fluxes than in the
LMC. Given the dominant contribution from low-mass stars, in this case
the treatment of the RGB mass loss is more relevant than for the LMC.
While the expected location of the peak is essentially independent of
the amount of mass lost by the stars while climbing along the RGB,
to reproduce the faint side of the luminosity function and to derive
an average J magnitude $\rm M_J \sim -6.18$ mag, in agreement with the
observations, it is necessary to assume that stars with mass in the
$\rm 1-1.3~M_{\odot}$ range lose $\rm 0.3~M_{\odot}$ before reaching
the TRGB.

\begin{acknowledgements}
CV and PV acknowledge support by the INAF-Theory-GRANT 2022 
“Understanding mass loss and dust production from evolved stars”.
\end{acknowledgements}

%
%

\end{document}